\documentclass[preprint]{aastex}
\newcommand{\beq}{\begin{equation}}
\newcommand{\eeq}{\end{equation}}
\def\stacksymbols #1#2#3#4{\def\theguybelow{#2}
        \def\verticalposition{\lower#3pt}
        \def\spacingwithinsymbol{\baselineskip0pt\lineskip#4pt}
        \mathrel{\mathpalette\intermediary#1}}
\def\intermediary #1#2{\verticalposition\vbox{\spacingwithinsymbol
        \everycr={}\tabskip0pt
        \halign{$\mathsurround0pt#1\hfil##\hfil$\crcr#2\crcr
                \theguybelow\crcr}}}

\catcode`\@=11 
\def\gsim{\ifmmode{\mathrel{\mathpalette\@versim>}}
    \else{$\mathrel{\mathpalette\@versim>}$}\fi}
\def\lsim{\ifmmode{\mathrel{\mathpalette\@versim<}}
    \else{$\mathrel{\mathpalette\@versim<}$}\fi}
\def\@versim#1#2{\lower 2.9truept \vbox{\baselineskip 0pt \lineskip 
    0.5truept \ialign{$\m@th#1\hfil##\hfil$\crcr#2\crcr\sim\crcr}}}
\catcode`\@=12
\def\bd{B2$_{02}\,$}
\def\bt{B3$_{02}\,$}
\def\lx{L_{\rm X}}
\def\lbh{L_{\rm BH}}
\def\tx{T_{\rm X}}
\def\ta{T_{\rm a}}
\def\tp{T_{\rm p}}
\def\Emis{{\cal{E}}}

\def\mbh{M_{\rm BH}}
\def\mdot{\dot\mbh}
\def\ledd{L_{\rm Edd}}
\def\re{R_{\rm e}}
\def\eps{\epsilon}

\def\epsw{\eps_{\rm w}}
\def\epswM{\epsw^{\rm M}}
\def\lbhefopt{L_{\rm BH,opt}^{\rm eff}}

\def\Msun{M_{\odot}}

\def\lb{L_{\rm B}}
\def\mast{M_*}

\def\mgas{M_{\rm gas}}

\lefthead{S. Pellegrini, L. Ciotti, J.P. Ostriker}
\righthead{X-ray signatures of AGN feedback in Es}

\begin{document}
\title{X-ray properties expected from AGN feedback in elliptical galaxies}

\author{Silvia Pellegrini\altaffilmark{1},
Luca Ciotti\altaffilmark{1}, 
Jeremiah P. Ostriker\altaffilmark{2,3}
}
\affil{$^1$Department of Astronomy, University of Bologna,
via Ranzani 1, I-40127, Bologna, Italy} 
\affil{$^2$Princeton University Observatory, Princeton, NJ 08544, USA}
\affil{$^3$IoA, Cambridge, UK}

\slugcomment{Submitted to ApJ}

\begin{abstract} 
The evolution of the interstellar medium (ISM) of elliptical galaxies
experiencing feedback from accretion onto a central supermassive black
hole has been studied recently with high-resolution 1D hydrodynamical
simulations; these included cooling, heating and radiative pressure
effects on the gas, specific for an average AGN spectral energy
distribution, a RIAF-like radiative efficiency, mechanical energy,
mass and momentum input from AGN winds, and the effects of starbursts
associated with accretion.  Here we focus on the observational
properties of the models in the soft and hard X-ray bands,
specifically on 1) the nuclear X-ray luminosity; 2) the global X-ray
luminosity and temperature of the hot ISM; 3) its temperature and
X-ray brightness profiles, during quiescence, and before, during and
after an outburst. After an evolution of $\sim 10$ Gyr, the bolometric
nuclear emission is very sub-Eddington ($l \sim 10^{-4}$), and within
the range observed, though larger than the most frequently observed
values.  The nuclear bursts last for $\approx 10^7$ yrs, and the
duty-cycle of nuclear activity is a few$\times (10^{-3}-10^{-2})$,
when calculated over the last 6 Gyr.  The ISM thermal luminosity $\lx$
oscillates in phase with the nuclear one, but drawing much broader
peaks; a comparison with observed $\lx$ values, for galaxies of
optical luminosity similar to that of the models, shows that this
behavior helps reproduce statistically the observed large $\lx$
variation. At the present epoch, the largest observed $\lx$ could be
reproduced only by adding an external confining medium.  The average
gas temperature is within the observed range; when limited to within
$\re$, its values lie on the upper half of those observed.  In
quiescence, the temperature profile has a negative gradient; thanks to
past outbursts, the brightness profile lacks the steep shape typical
of inflowing models. After outbursts, disturbances are predicted in
the temperature and brightness profiles (as analyzed also with the
unsharp masking).  Most significantly, during major accretion
episodes, a hot bubble from shocked hot gas is inflated at the galaxy
center (within $\approx 100$ pc); the bubble would be conical in
shape, in real galaxies, and show radio emission. Its detection in the X-rays is within
current capabilities, though it would likely remain unresolved.  The
ISM resumes its smooth appearance on a time-scale of $\approx 200$
Myr; the duty-cycle of perturbances in the ISM is of the order of
5-10\%.  The present analysis reveals an agreement of the models with
the observations, but also evidences that additional input physics is
important in the ISM-black hole coevolution, to fully account for the
properties of real galaxies. The main insertions to the models would
be a confining external medium and a jet.  The jet will reduce further
the mass available for accretion (and then $l$), and may help,
together with an external pressure, to produce flat or positive
temperature gradient profiles (that are common among galaxies in high
density environments).  Alternatively to the jet, a reduction of $l$
can be obtained if the switch from high to low radiative efficiency
takes place at a larger $l$ ($\simeq 0.1$) than routinely assumed
($\simeq 0.01$).
\end{abstract}


\section{Introduction}
\label{intro}

Supermassive black holes (MBHs) at the centers of bulges and
elliptical galaxies play an important role in the processes of galaxy
formation and evolution (e.g., Cattaneo et al.~2009), as testified by
remarkable correlations between host galaxy properties and the MBH
masses (e.g., Magorrian et al.~1998, Ferrarese \& Merritt 2000,
Gebhardt et al.~2000, Graham et al.~2001) and as supported by many
theoretical studies (e.g., Silk \& Rees 1998, Haiman, Ciotti \&
Ostriker 2004; Merloni et al.~2004, Sazonov et al.~2005, Di Matteo,
Springel \& Hernquist 2005, Hopkins et al.~2006, Somerville et
al.~2008, Kormendy et al.~2009).  An important aspect of the
coevolution process is the radiative and mechanical feedback by the
accreting MBH onto the galactic interstellar mediun (ISM) that is
continuously replenished by normal stellar mass losses, at a rate of
the order of $\approx 1 M_{\odot}$ yr$^{-1}$ in a medium-mass galaxy.
In absence of feedback from a central MBH (and stripping from the 
intracluster medium
in case of satellite galaxies), this ISM would develop a flow directed
towards the galactic center, accreting $\gsim 1 M_{\odot}$ yr$^{-1}$
in a process similar to a ``cooling flow''.  Instead, in low mass
galaxies, type Ia supernova (SNIa) heating is able to sustain a
low-luminosity, global galactic wind (e.g., Ciotti et al.~1991, David
et al.~1991, Pellegrini \& Ciotti 1998), and the central MBH is in a
state of permanent, highly sub-Eddington hot accretion (Ciotti \&
Ostriker 2011).

Therefore, in medium to high mass galaxies, feedback is required by
the following empirical arguments: 1) the large amount of gas lost by
the passively evolving stellar population during the galaxies'
lifetime is not observed (e.g., Peterson \& Fabian 2006), and just
$\lsim $1\% of the mass made available by stars is contained in the
masses of present epoch MBHs; 2) bright AGNs, as would be expected
given the predicted mass accretion rate, are not commonly seen in the
spheroids of the local Universe (e.g., Fabian \& Canizares 1988,
Pellegrini 2005). Thus AGN feedback is required just on the
basis of a mass balance argument. Since over a large part of the
galaxies extent, and for a large fraction of their lifetime, the ISM cooling
time $t_{cool}$ is much lower than the galaxy age, one early quasar
phase cannot be the solution to the cooling flow problem. The solution
requires either steady heating, or heating with bursts on a timescale
$\Delta t\approx t_{cool}$.  Unfortunately, on a purely theoretical
ground, how much radiative and mechanical energy and momentum output
from the MBH can effectively interact with the surrounding ISM, and
what are the resultant MBH masses, is difficult to establish.

Recently, the interaction of the MBH output with the inflowing gas has
been studied with high-resolution 1D hydrodynamical simulations in a
series of papers (Ciotti \& Ostriker 2007; Ciotti, Ostriker \& Proga
2009, 2010 hereafter Papers I and III; Shin, Ostriker \& Ciotti
2010a,b; Ostriker et al. 2010; Jiang et al.~2010), that are
currently being extended to 2D treatments (Novak, Ostriker \& Ciotti
2010).  These simulations implement a physically based detailed
treatment of the radiative energy and momentum input from the MBH into
the ISM, consistent both with observed average AGN spectra and
theoretical calculations of radiation transport; they also include
starformation, and a modelling of the mechanical energy and momentum
feedback from AGN winds. The combined effects of radiative and
mechanical feedback produce recurrent AGN burst phases accompanied by
starformation, spaced apart by longer phases of relative quiescence. A
cycle repeats with the galaxy seen alternately as an AGN/starburst for
a small fraction of the time, and as a ``normal'' elliptical for much
longer intervals. Accretion fueled feedback thus proves effective in
suppressing long lasting cooling flows and in maintaining MBH masses
within the range observed today, since the gas is mostly lost in outflows
or consumed in starbursts.  Remarkably, {\it while star formation is
  suppressed when the AGN is in the low-luminosity state, it is
  enhanced by the strong AGN outbursts, consistent with observations}
(Schawinski et al. 2009).  Note finally that  a major role in
producing global degassing, and in regulating the flow
evolution, is also played by the SNIa's heating.

The previous papers, with the exception of Pellegrini, Ciotti \&
Ostriker (2009), were mainly dedicated to the study of the accretion
physics and the feedback effects. Here instead we focus on the
appearance that the models would have if observed in the X-ray band,
both in quiescence and during an outburst of activity. We concentrate
on two models (named \bd and \bt) extracted from the suite of cases
presented in Paper III (Table 1 therein), characterized by a
mechanical feedback efficiency dependent on the Eddington scaled
accretion luminosity. \bd and \bt were considered particularly
successful, since their input parameters agree with previous
theoretical studies or observations (as, e.g., for the AGN wind
opening angle, and the peak value of mechanical efficiency, that are
in accord with those estimated from 2D and 3D numerical simulations;
Proga, Stone \& Kallman 2000, Proga \& Kallman 2004, Benson \& Babul
2009), and, at the same time, their final properties (as mass fraction
of a younger stellar population, MBH mass, etc.) are in reasonable
accord with observations.  The only difference in the input physics of
\bd and \bt is the maximum value of the mechanical efficiency.  Since
in the simulations the treatment of feedback is physically based, not
tuned to reproduce observations, any agreement or discrepancy of the
resulting model properties with X-ray observations is relevant to
improving our understanding of the MBH-ISM coevolution, putting
further constraints on the input ingredients, and possibly telling us
what additional physics may be important in the problem. However, we
stress that the models describe an isolated galaxy, where ram-pressure
stripping (in case of satellite galaxies) and intracluster medium
pressure confinement (in case of group or cluster central galaxies)
are not taken into account (see Shin et al.~2010b).

The main observational signatures investigated here include the
nuclear and gaseous emissions, and the ISM temperature and brightness
profiles in the quiescent phases, and before, during, after a
burst. Particular attention is paid to the appearance and
detectability of central hot bubbles, with diameters of $\sim$a
hundred parsecs, that are produced by the models during outbursts, and
to various kinds of disturbances in the hot ISM.  The analysis is
performed in the soft and hard X-ray bands, also after
unsharp-masking.  These predictions are relevant for their
observational consequences, since the high angular resolution of the
$Chandra$ satellite has allowed us to obtain the best definition ever
for the hot gas properties of the galaxies of the local universe, by
separating the contributions of stellar sources and hot gas, and the
emission coming from different spatial regions within galaxies (e.g.,
Fabbiano 2011).  In particular, in
several elliptical galaxies various kinds of hot gas disturbances have
been detected, likely resultant from nuclear activity (e.g.,
Finoguenov \& Jones 2001, Jones et al.~2002, Forman et al.~2005,
Machacek et al.~2006, O'Sullivan et al.~2007, Million et al. 2010).
At the same time, nuclear emission values have been detected down to
very low luminosities, comparable to those of X-ray binaries
(e.g., Loewenstein et al.~2001, Gallo et al. 2010, Pellegrini 2010).

The paper is organized as follows. Section 2 summarizes the main
evolutionary phases of the representative models (\bt and \bt)
considered; Section 3 describes how the observational properties of
the models are derived; Section 4 presents a comparison of the nuclear
luminosities with existing observations; Section 5 discusses the
evolution of the X-ray luminosity and emission-weighted temperature of
the ISM; Section 6 presents the projected temperature and surface
brightness profiles at representative times during quiescence and
a nuclear outburst. Finally, in Section 7 we summarize and
discuss the main results.

\section{Two representative models: main features}
\label{repmod}

The basic ideas behind the present class of models for feedback
modulated accretion flows have been
introduced in Ciotti \& Ostriker (1997, 2001), and Ostriker \& Ciotti
(2005), and developed in detail in the papers listed in the
Introduction; a comprehensive recent discussion is given in
Ciotti \& Ostriker (2011).  Here we focus on 
models \bd and \bt from Paper III
(where the description of the numerical code and the input
physics is given).  The initial parameter values and the main final
properties of the models are given in Table~\ref{modref}. 

The two models refer to an isolated elliptical galaxy
placed on the Fundamental Plane, with a projected central stellar
velocity dispersion $\sigma =260$ km s$^{-1}$, a total B-band
luminosity $\lb=5\times 10^{10}L_{\rm B\odot}$, and an effective
radius $\re=6.9$ kpc.  The stellar density profile is described by a
Jaffe (1983) law, and the dark halo profile is such that the total
(stellar+dark) mass density profile scales as $\rho \propto r^{-2}$ at
large radii; all relevant dynamical properties used in the code are
discussed in Ciotti, Morganti \& de Zeeuw (2009). The dark-to-visible
mass ratio is one within $\re$, and the resulting stellar
mass-to-light ratio is $M_*/\lb=5.8$. Finally, a standard SNIa's rate
declining with time $t$ as $t^{-1.1}$ is assumed.  The initial MBH mass is
set to $10^{-3}$ the initial stellar mass $M_*$, i.e., it is $2.9\times
10^8\Msun$.  The simulations begin at a galaxy age of $\sim 2$ Gyr
(that is a redshift $z\sim 2$, the exact value depending on the epoch
of elliptical galaxy formation, usually put at $z\gsim 2$).

The efficiency for producing radiation (Soltan 1982, Yu \& Tremaine 2002)
of material accreting on the MBH at the rate $\mdot$ is 
\begin{equation}
\epsilon =0.2\times {A \dot m \over 1+A\dot m }, 
\label{effic}
\end{equation}
where $\dot m=\mdot/\dot M_{\rm Edd}$ is the Eddington-scaled
accretion rate. Thus, for $A=100$, one has $\epsilon \sim 0.2$ at
large mass accretion rates $\dot m
\gg 0.01$, and $\epsilon$ declining as for radiatively inefficient accretion
flows (RIAFs, Narayan \& Yi 1994), as  $\epsilon \sim 20 \dot
m$, for $\dot m\lsim 0.01$ (see also Sect.~\ref{nuc} for additional
comments on this choice).  The mechanical feedback implemented is
that of the Broad Line Region winds (leading to outflow velocities of
$\simeq 10^4$ km/s, similar to what observed, e.g. Chartas et
al.~2003, Crenshaw et al.~2003), and reproduces the main
features of the numerical modeling by Proga (2003). In particular, the
mechanical efficiency scales with the Eddington ratio
$l=\lbh/\ledd$ (where $\lbh =\epsilon\mdot c^2$ is the instantaneous
bolometric accretion luminosity), reaching a maximum value $\epswM$ of $3\times
10^{-4}$ (for \bt), and $10^{-3}$ (for \bd), when $l\gsim
2$; also, the aperture solid angle of the conical nuclear wind
increases at increasing $l$. Note that the values of the mechanical
efficiency in cols. (2) and (3) of Tab.~\ref{modref} are to be
contrasted with the generally higher fixed efficiency of $5\times 10^{-3}$ 
commonly adopted in the literature (e.g., Hopkins et al. 2005, Di Matteo et al. 2005,
Johansson et al. 2009). The mechanical output of a nuclear jet is
also computed, but not added to the hydrodynamical equations, and it
will be inserted in a future work.

The evolution of the gas flows is obtained integrating the
time-dependent (1D) Eulerian equations of hydrodynamics, with a
logarithmically spaced and staggered radial grid, extending from 2.5
pc from the central MBH to 250 kpc. It is most important that the 
resolution is high enough that the inner boundary is within the
Bondi radius (Bondi 1952); if this is not ensured, the accretion rate will be
calculated incorrectly. Thus ``Bondi accretion'' is not assumed; the
correct, time-dependent accretion rate is computed from the
hydrodynamical equations.

The code derives self-consistenly the source and sink terms of mass,
momentum and energy associated with the evolving stellar population
(stellar mass losses, SNIa events), the temporal steepening of the
stellar velocity dispersion within the sphere of influence of the MBH
as a consequence of its growth, the star formation during nuclear
starbursts, and finally accretion and MBH feedback.  Needless to say,
the code conserves mass, energy, and momentum (e.g., Ostriker et
al. 2010).  Gas heating and cooling are calculated for a photoionized
plasma in equilibrium with an average quasar spectral energy
distribution (as detailed by Sazonov et al.~2005), and the resulting
radiation pressure and absorption/emission are computed and
distributed over the ISM from numerical integration of the radiative
transport equation. The effects of radiation pressure on dust due to
the starburst luminosity in the optical, UV and Infrared are also
considered. A circumnuclear accretion disk is modeled at the level of
``sub-grid'' physics, and a set of differential equations describing
the mass flow on the disk, its star formation rate, mass ejection and
finally MBH accretion are solved at each time-step.

\begin{figure}
\hskip -0.8truecm
\includegraphics[angle=0.,scale=0.55]{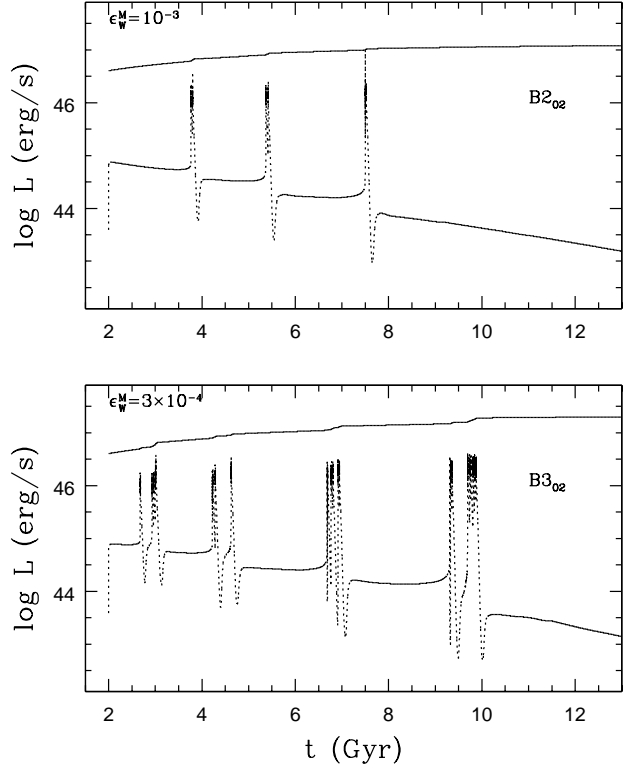}
\caption{Time evolution of the nuclear luminosity for models \bd (upper
 panel) and \bt (bottom panel).  The Eddington luminosity $\ledd$
 (the almost horizontal solid line), and the bolometric luminosity
 resulting from accretion on the MBH, $\lbh =\epsilon\mdot c^2$
 (dotted line), are shown.  The larger number of bursts shown by \bt, their
 larger temporal extension and substructure, are due to
 the reduced peak value of mechanical efficiency $\epswM$ of the AGN
 wind (see Sect.~\ref{repmod}).
The bursts become rarer for increasing time, in pace
 with the decreasing mass return rate from the evolving stellar
 population.}
\label{f1}
\end{figure}

The resulting evolution of $\lbh$ for \bd and \bt is shown
in Fig.~\ref{f1}. These models have fairly standard values
of the input parameters, and their general properties in the X-ray
band, are typical of the class of 1D models investigated in Paper III.
At the beginning, the galaxy is replenished by gas produced by the 
mass return from the evolving stars.
Soon AGN outbursts develop, due to accretion of this gas,
accompanied by star formation, and followed by degassing and 
a precipitous drop of the nuclear accretion rate.  The outbursts are
separated by long time intervals during which the galaxy is
replenished again by gas from the stellar mass losses.  

The behavior of the gas during an outburst is almost independent of
the specific burst episode considered.  The outburst precursor is the
off-center growth of a thin shell of dense gas (at a radius of $\sim
0.5-1$ kpc) that progressively cools below X-ray emitting temperatures
and falls towards the center; compression of the enclosed gas follows
and a central burst is triggered, even before the cold shell reaches
the center\footnote{In the code the gas is allowed to cool down to 
a minimum temperature of $10^4$ K.}.
In less than a million years, a radiative shock
originates from the center and quickly (in $\approx 10^6$ yrs)
produces an outward moving shell that collides with the original shell
falling in. Reflected shock waves carry fresh material for accretion,
and produce new sub-bursts. This leads to the rich temporal structure
of each outburst event, especially visible in model \bt (bottom panel,
Fig.~\ref{f1}). Eventually, the cold material left after starformation
(Ciotti \& Ostriker 2007) is accreted in a final
burst, a major shock leaves behind a very hot and dense center, and
causes a substantial galaxy degassing.  In general, while radiative
effects mainly work on the kpc scale, mechanical feedback from the AGN
winds is more concentrated and affects the ISM on the $\sim 100$ pc
scale (see Fig.~11 in Paper III).  During the next few tens of million
years, the gas cools, resumes its subsonic velocity, the density
starts increasing again due to stellar mass losses, and the cycle
repeats.  At late epochs, the gas flows finally set in a state of
steady, hot and low-luminosity accretion.

\section{Observational properties of the models}
\label{obsprop}

The observational model properties considered in Paper III were the
nuclear bolometric, optical and UV luminosities, and the X-ray
luminosity of the ISM within $10\re$.  The latter was evaluated
fiducially just by cutting the bolometric gas emission below a
threshold temperature of $5\times 10^6$ K.  In the following, we
calculate the X-ray luminosity of the nucleus ($L_{\rm BH,X}$,
Sect.~\ref{nuc}), and, with a more detailed and realistic procedure,
the total luminosity and emission weighted temperature of the hot gas
($\lx$ and $<\tx >$, Sect.~\ref{global}); we also calculate the
temperature and the surface brightness profiles during quiescent
times, and during an outburst (Sect.~\ref{proj}).  We briefly describe
below how the observational gas properties are calculated from the
numerical outputs for the gas density and temperature.

The X-ray emission of the different model components is calculated
over the energy range of 0.3-8 keV (the $Chandra$
sensitivity band), and also in two separate bands, 0.3-2 keV
(``soft'') and 2-8 keV (``hard''), typically used in studies of the
nuclear and gaseous properties. In practice, at any given time, the
volume gas luminosity is calculated from the gas density and
temperature distribution on the numerical grid as
\begin{equation}
\lx=4\pi \int_0^{\infty}\Emis (r) r^2 dr,
\label{lumeq}
\end{equation}
where the emissivity is given by $\Emis(r) = n_{\rm e}(r) n_{\rm H}(r)
\Lambda [T(r), Z]$, $n_{\rm e}$ and $n_{\rm H}$ are the number
densities of electrons and hydrogen, and $\Lambda (T,Z)$ is the
cooling function. The cooling function is calculated over the two
energy intervals by means of the radiative emission code APEC, valid
for hot plasmas at the collisional ionization equilibrium (Smith et
al.~2001), as available in the XSPEC package for the analysis of the
X-ray data. For simplicity we choose constant abundance at the
solar value, and the solar
abundance ratios of Grevesse \& Sauval (1998), which is consistent
with observed gas metallicities (e.g., Loewenstein \& Davis 2010).  In
order to speed-up the analysis, we derived with APEC the values of
$\Lambda$, for each energy band, for a large set of temperatures in
the range 0.1-16 keV; then we obtained two very accurate non-linear
fits of the tabulated values (with maximum deviations $<1$\%, see
Ciotti \& Pellegrini 2008). These fits were used to compute the
integral in (\ref{lumeq}), and in every other integration where the
emissivity is needed.  For example the emission weighted
  temperature for the whole galactic volume was computed as
\begin{equation}
<\tx> = {4\pi\over\lx}\int_0^{\infty}\Emis (r) T(r) r^2 dr.
\label{teq} 
\end{equation}
The surface brightness profile $\Sigma (R)$, the emission weighted projected
temperature profile $\tp(R)$, and the associated emission weighted aperture
temperature profile $\ta (R) $, were obtained by numerical
integration of the simulation outputs as
\begin{equation}
\Sigma (R)=2\int_R^{\infty}{\Emis(r) r\over\sqrt{r^2-R^2}}dr,
\label{sig}
\end{equation} 
\begin{equation}
\tp(R) = {2\over \Sigma (R)} 
\int_R^{\infty}{T(r)\Emis(r) r\over\sqrt{r^2-R^2}}dr,
\label{tprof} 
\end{equation}
\begin{equation}
\ta (R) ={\int_0^R \tp (R')\Sigma(R')R' dR' \over\int_0^R \Sigma(R') R' dR' }.
\label{ta} 
\end{equation}
The accuracy of the integrations above is verified by checking that
the surface integral of $\Sigma (R)$ over the whole grid
recovers the same luminosity calculated via equation (\ref{lumeq}),
and that $\ta(\infty)=<\tx>$ within few percent (Ciotti \& Pellegrini
2008).  The surface integral of $\Sigma (R)$ is also used to compute
the gas emission within the optical effective radius $R_{\rm e}$,
and in equation (\ref{ta}) to compute the average temperature within
the optical effective radius.

In order to highlight local departures from the mean ISM brightness
profile, and to evidence major brightness features that could be
revealed by observations, ``fluctuation'' profiles have been also
created. These have been constructed with a technique similar to
the so-called  unsharp masking, frequently adopted in
observational analysis (e.g., Fabian et al.~2003). In practice, the
brightness profiles $\Sigma (R)$ have been convolved with a 2D Gaussian of
dispersion $\sigma$:
\begin{equation}
{\rm PSF}={e^{-{R^2\over 2\sigma^2}}\over 2\pi\sigma^2},
\label{psf} 
\end{equation}
so that the resulting surface brightness profile can be written as
\begin{equation}
\Sigma_{\rm obs}(R)=\int_0^{\infty}
{\rm I}_0\left({RR'\over\sigma^2}\right) \,\,  {e^{-{R'^2+R^2\over2\sigma^2}}
  \over \sigma^2} \,\, \Sigma (R')R'dR',
\label{conv} 
\end{equation}
where I$_0$ is the zeroth-order modified Bessel function of first
kind. In the analysis of the simulations, the
integral above is solved numerically, after a careful choice of
$\sigma$.
As expected, a too large $\sigma$
produces an almost featureless profile, while a too small $\sigma $
reproduces the unprocessed profile. After some attempts, it turned out that,
in order to highlight local features, the optimal choice is that of
a $\sigma $ equal, at each gridpoint, to the sum of the lengths of
the immediately preceding and subsequent grid intervals.
 The ``unsharp-masked'' profile is then defined in a natural way as
\begin{equation}
\Sigma_{\rm UM}(R)\equiv {\Sigma (R)\over \Sigma_{\rm obs}(R)}-1.
\label{conv} 
\end{equation}

\section{Nuclear luminosities}
\label{nuc}

Figure~\ref{f1} shows the time evolution of the nuclear bolometric
accretion luminosity $\lbh$ for \bd and \bt (whose input parameters
differ only for the maximum value of the mechanical efficiency
$\epsilon_{\rm w}^{\rm M}$, that is respectively $10^{-3}$ and
$3\times 10^{-4}$, Tab.~\ref{modref}). Strong intermittencies at an
earlier epoch, with $\lbh$ reaching the Eddington value, become rarer
and rarer with time, as the mass return rate from the stellar
population declines, until a smooth, hot, and very sub-Eddington
accretion phase establishes. The different mechanical efficiency is
responsible for the sharp bursts in model \bd, and the more
time-extended and structured bursts in model \bt (Paper III).  Towards
the present epoch, at a galaxy age of 12 Gyr, the mass accretion rate
on the MBH for both models is $\mdot\approx 0.02\, \Msun$/yr, that
translates into an Eddington scaled accretion rate $\dot m \simeq
1.7\times 10^{-3}$ and $\dot m \simeq 1.2\times 10^{-3}$ respectively
for \bd and \bt.  The value of $\dot m$ of \bt is a bit
smaller than for \bd, because of its larger final $\mbh$
(Tab.~\ref{tab2}), a consequence of the weaker mechanical feedback.
At the present time, and during the interburst periods, accretion has
then entered the RIAF regime, and the radiative efficiency is
$\epsilon \simeq 0.02$; the nuclear bolometric luminosity is
$\lbh=2.4\times 10^{43}$ erg s$^{-1}$ for both models, and the
corresponding Eddington ratios are $l\simeq 2\times 10^{-4}$ (\bd) and
$l\simeq 10^{-4}$ (\bt), see Tab.~\ref{tab2}.

These results agree with the observation that in the local universe
massive MBHs are mostly radiatively quiescent, and the fraction of
them at luminosities approaching their Eddington limit is negligible
(e.g., Ho 2008).  For example, in the sample of nuclei of the Palomar
Spectroscopic Survey of northern galaxies, a nearly complete sample,
magnitude limited at $B_{\rm T} \leq 12.5$ mag, $\sim 50$\% of
ellipticals show detectable emission line nuclei\footnote{A similar
  fraction ($\sim 52$\%) of the sample of red sequence galaxies of the
  SDSS ($r<17.77$, median redshift $z=0.1$) have detectable line
  emission (Yan et al.~2006).}, but mostly of low level ($L_{\rm
  H\alpha}<10^{40}$ erg/s; Ho 2008).  For this sample, the nuclear
bolometric luminosity $(L_{\rm bol,nuc})$ was derived from the
observed nuclear 2-10 keV emission, using the correction $L_{\rm
  bol,nuc}/L_{\rm X,nuc}=15.8$ (Ho 2009). It was found that elliptical
galaxies span a large range of $L_{\rm bol,nuc}$, from $10^{38}$ to
$10^{43}$ erg s$^{-1}$, with a median value of $L_{\rm bol,nuc}\simeq
1.7\times 10^{40}$ erg s$^{-1}$ and a mean value of $4.6\times
10^{41}$ erg s$^{-1}$; the Eddington scaled luminosity $l$ has a
median value of $l=1.2\times 10^{-6}$ (mean $l=1.2\times 10^{-5}$),
with $l< 10^{-3}$ for elliptical galaxies, and extending down to
$l=10^{-8}$. The representative models tend therefore to lie on the
upper end of the observed distributions of $L_{\rm bol,nuc}$ and $l$,
a result that probably remains true regardless of the uncertainty in
the bolometric correction from the 2-10 keV band. We will return on
this point (already noticed in Papers I and III) in the Conclusions.

Another way of comparing the model $\lbh$ with observed values, is to
estimate $L_{\rm BH,X}$ of the models, and compare it directly with
observed $L_{\rm X,nuc}$ values.  Accurate $L_{\rm X,nuc}$
measurements in a large number have been derived recently for
elliptical galaxies of the local universe, based on $Chandra$ data
(e.g., Gallo et al.~2010, Pellegrini 2010); in the 0.3--10 keV band,
$L_{\rm X,nuc}$ ranges from $\gsim 10^{38}$ erg s$^{-1}$ to $\sim
10^{42}$ erg s$^{-1}$, with most of $L_{\rm X,nuc}/\ledd$ as low as
$10^{-6}$--$10^{-8}$.  In particular, for the final MBH masses of the
models (Tab.~\ref{tab2}), it is observed that $10^{38}\lsim L_{\rm
  X,nuc}$(erg s$^{-1})\lsim 10^{42}$, and $10^{-9}\lsim L_{\rm
  X,nuc}/\ledd \lsim 10^{-4}$, both from the sample of the $Hubble$
Virgo Cluster Survey (Gallo et al.~2010), and that of 112 early type
galaxies within $\sim 60$ Mpc (Pellegrini 2010).  The 0.3-10 keV
nuclear luminosity $L_{\rm BH,X}$ of the models at the present epoch
can be derived adopting a correction factor appropriate for the
spectral energy distribution of a radiatively inefficient accretion
flow, i.e., $L_{\rm BH,X} \lsim 0.2\lbh$ for low luminosity AGNs
(Mahadevan 1997). This gives $L_{\rm BH,X} \lsim 5\times 10^{42}$ erg
s$^{-1}$, and $L_{\rm BH,X}/\ledd\lsim 2\times 10^{-5}$.  Also in the
X-ray band, then, the model nuclear luminosity tends to be larger than
what typically observed; $L_{\rm BH,X}/\ledd$ is within the observed
range, though lying in its upper end.  All this may suggest that in real
galaxies an additional mechanism is reducing further the mass
available for accretion, as could be provided by a nuclear jet, and/or
a thermally driven wind from a RIAF (Blandford \& Begelman 1999).
In the latter case, only a
fraction of the gas within the accretion radius actually reaches the
MBH; the binding energy released by the accreting gas is transported
radially outward to drive away the remainder in the form of a
wind. 

Alternatively, it may be that the quadratic dependence of  $l$
on $\dot m$ (Sect.~\ref{repmod}) sets in at a higher value of $\dot m$
than adopted in Eq.~\ref{effic}, for which it starts at $\dot m\sim 10^{-2}$.
Within the uncertainties on the observational and theoretical input,
we might have chosen the constant $A=10$ 
rather than 100; in this way, the quadratic dependence
would have set in at $l \lsim 0.1$ rather than $l\lsim 0.01$. This would have
reduced the late time nuclear luminosity by a factor of $\simeq 9$, as
confirmed by a supplementary run\footnote{The model evolution of
course also changes, because of the smaller direct radiative feedback, and the smaller indirect
 mechanical feedback, that in the models is a function of $\lbh$.
The bursts are more extended in time, with a
larger accreted MBH mass, which reduces slightly
also the final $l$ ratio.} of models \bd and \bt with $A=10$.

Finally, another interesting - albeit more difficult - comparison with
observational results can be done using the duty-cycle. The latter can
be calculated as the fraction of the total time spent by the AGN in
the ``on'' state, defined by a luminosity greater than $\ledd/30$ in a
given band, over some temporal baseline (Paper
III)\footnote{Alternatively, the duty-cycle can be obtained as a
  luminosity weighted average over a chosen  time interval,
  with very similar results.}.  So doing, the duty-cycle turns out to
be a small number (Tab.~\ref{tab2}); for example, for a temporal
baseline ending at the present epoch (i.e., at 13 Gyr in
Fig.~\ref{f1}), the duty-cycle of model \bd is zero when starting from
9 Gyr (at a redshift $z\approx 0.45$, for a flat universe with
$H_0=71$ km s$^{-1}$ Mpc$^{-1}$, $\Omega_M=0.27$,
$\Omega_{\Lambda}=0.73$), as no burst occurs after $\simeq 7.5$ Gyr;
when starting from 6 Gyr ($z\approx 1$), it is $\simeq 6.3\times
10^{-3}$, $3.2\times 10^{-3}$, and $3.0\times 10^{-3}$ respectively in
the bolometric, optical (absorbed), and UV (absorbed) bands. For model
\bt, in the same bands, the duty-cycle is respectively $\simeq
4.9\times 10^{-2}, 1.7\times 10^{-2}, 7.1\times 10^{-3}$ (when
starting at 9 Gyr), and $\simeq 4.8\times 10^{-2}, 1.8\times 10^{-2},
8.6\times 10^{-3}$ (when starting at 6 Gyr).  The duty-cycle decreases
from the bolometric, to the optical, to the UV bands, due to the
different values of the opacity in these bands.

These duty-cycle values are broadly consistent with the fraction of
active galaxies measured in observational works.  For example, Greene
\& Ho (2007) estimated the (mass dependent) number of active galaxies,
using broad-line luminosities from SDSS DR4, for galaxies with $z <
0.352$ (age of the universe $\gsim 10$ Gyr); statistically speaking,
the fraction of active systems can be interpreted as a duty cycle for
MBHs in a given mass range. Greene \& Ho report duty-cycle values of
the order of $4\times 10^{-3}$ for $10^7$ $M_{\odot}$ MBHs, declining
at increasing mass. Similar duty-cycle values of $\sim 2\times
10^{-3}$, decreasing at increasing MBH mass, are reported by Heckman
et al. (2004).  The duty-cyles of the models tend to be larger than
those observed; however, the comparison is limited by the small number
of models considered, and the fact that the only way to compute
duty-cycles different from zero is to extend the temporal baseline
back in the past.  A more consistent comparison with observations can
be made with an increased dataset of models, and computing the duty
cycle for the last 2--3 Gyrs. Clearly this procedure will reduce the
duty cycle, as the models are characterized by a declining nuclear
activity. Again, the computed duty cycle would have been reduced
significantly also if we had raised the threshold for RIAF-like
behavior of the radiative efficiency to $l=0.1$.

\section{Luminosity and Temperature of the Gas}
\label{global}

We describe here the time evolution of the global thermal
luminosity and temperature of the ISM.

\subsection{Luminosity evolution}
\label{lum}

The top panels of Fig.~\ref{f2} show the time evolution of the total
gas emission $\lx$ for models \bd (left) and \bt (right).  Red lines
refer to the soft (0.3-2 keV) band, and blue lines to the hard (2-8
keV) band; for reference, the scaled-down $\lbh$
(black dotted line) is also shown.

The luminosity evolution of the gas for the two models is
qualitatively similar, with peaks in $\lx$ coinciding with those in
$\lbh$.  The sudden increase of the gas emission during outbursts is
due to the increase in temperature and density in the central galactic
regions ($\simeq 10^2-10^3$ pc), caused by radiative gas heating
(Compton and photoionization), and by compression due to direct and
reflected shock waves, produced by mechanical and radiative feedback,
that are associated with the AGN and the starburst.  For short times,
most of the luminosity in the peaks of $\lx$ originates from a very
small region at the galactic center ($\approx 100$ pc), thus it is
observationally indistinguishable from the luminosity of the nucleus
(see also Sect.~\ref{brilx}).  The hard emission oscillates in phase
with the soft one, and presents the same overall behavior, but keeps
at a level almost 2 orders of magnitude lower.  Hard emission during
the outburst, as shown in Fig.~\ref{f2}, would be detectable with
$Chandra$, if centrally concentrated (see also Sect.~\ref{brilx}
below).  However, hard emission during quiescent times
would be difficult to distinguish from the contribution of unresolved
binaries, even with $Chandra$, if extended (e.g., Pellegrini et
al.~2007, Trinchieri et al.~2008).

A comparison of the peaks in $\lx$ and $\lbh$ reveals differences and
analogies.  While $\lbh$ shows sharp and sudden spikes at the
outbursts (increasing by 2 or more orders of magnitude in $\approx
10^{6-7}$ yr), and is almost constant between them, $\lx$ increases
slowly but steadily between outbursts, when the galaxy is replenished
by the stellar mass losses.  The peaks in $\lx$ become broader with
increasing time, but not weaker; for example, the increase of $\lx$
during the last major outburst of \bd is the largest one, with the
largest amount of gas heated and then removed from the galaxy in its
entire life.  Instead, when the burst ends, $\lx$ has the same abrupt
decrease as $\lbh$, due to the density drop following the final (and
usually strongest) sub-burst in each major accretion event.  Another
similarity is that both $\lbh$ and $\lx$ show sharper and ``cleaner''
bursts in \bd than in \bt: more radiatively dominated feedback bursts
(as in \bt) are richer in temporal substructure, because it takes
longer for the cold shells to be destroyed, thus more star formation
and MBH accretion occurs (Paper III).

The quiescent values of $\lx$ during the past few Gyr remain at the same level
of $\sim 10^{40}$ erg s$^{-1}$ for model \bd, and decrease by a small
factor of $\lsim 2$ to reach a present epoch $\lx=2\times 10^{39}$ erg
s$^{-1}$ for \bt. Previous compilations of observed
$\lx$ values (Fabbiano et al. 1992, O'Sullivan et al. 2001, Diehl \&
Statler 2007, Mulchaey \& Jeltema 2010) for early type galaxies of the
local universe, residing in all kinds of environments (from the field
to groups to clusters as Virgo and Fornax), show a range of $\lx$ from
$10^{40}$ erg s$^{-1}$ up to $10^{43}$ erg s$^{-1}$, at a B-band
optical luminosity of $\lb =5\times 10^{10}L_{\rm B,\odot}$ as the
model galaxy.  $\lx$ values larger than a few$\times 10^{41}$ erg
s$^{-1}$ belong to galaxies at the center of groups, or clusters, or
subclusters, for which an important contribution from the intragroup
or intracluster medium, or a confining action, is likely (e.g.,
Renzini et al. 1993, Brighenti \& Mathews 1998, Brown \& Bregman 2000,
Helsdon et al. 2001). However, the final $\lx$ of \bd and \bt is on the
lower end of the range of observed values; this indicates that
degassing is important in the models, and for many real cases it must
be impeded. In the simulations this could be obtained for example by
adding the external pressure from an outer medium (e.g., Vedder et
al.~1988), and it will be considered in future works.

\subsubsection{$\lx-\lb$ and $\lx-\lbh$ }\label{relaz}

Real galaxies
show a wide range of $\lx$ values, and 
the observed $\lx$ variation
has remained a subject of debate for years (e.g., Fabbiano et
al. 1989, Ciotti et al. 1991, White \& Sarazin 1991; Pellegrini 2011).
Thus, we check here whether the $\lx$ variation in the models 
during their evolution can
(partly) account for the large observed one. For this check we
considered the range of hot gas emission for the largest
sample of early type galaxies of the local universe (O'Sullivan et
al. 2001), after having excluded AGN-dominated cases, and central
dominant cluster or group galaxies.  Only galaxies with optical
luminosites within a range close to that of the model galaxy have been
considered (i.e., from log$L_B=10.5$ to log$L_B=10.8$, for which there
are 43 galaxies).  The discrete stellar sources contribution estimated
by O'Sullivan et al.  (2001) has been removed. The distribution of the
observed X-ray emission values so obtained is then compared with
that built for the soft X-ray emission of the models,
during the epoch from 2 to 12 Gyrs (Fig.~\ref{his}). Each emission
value enters the histogram with the fraction of the chosen epoch
during which it is present; the hypothesis
underlying this comparison is that statistically an observed galaxy
can be catched in anyone of the different phases shown in the past
2--12 Gyr by the representative models.

Figure~\ref{his} shows that $\lx$ of the models keeps within the
observed range, but it does not exceed significantly $\sim 10^{41}$
erg s$^{-1}$, while a fraction of galaxies populates this region.
Model \bd has a $\sim $constant $L_{X}$ in the past few Gyrs, so that
it populates mostly a couple of bins; model \bt (that experiences more
outbursts) reaches a wider coverage of the observed $L_X$ range, but
its $L_X$ distribution extends more to lower $L_X$ values than to
larger ones, due to a larger overall degassing.  Larger $\lx$ for the
models can be obtained when considering that real galaxies of similar
$\lb$ can have different values of the central stellar velocity
dispersion and effective radius.  These differences determine a
variety of flow evolution, and then of $\lx$ (Ciotti \& Ostriker
2011). Following this idea, Fig.~\ref{his} (bottom left) shows the
histogram of one possible variation to model \bt, that with
$\sigma=280$ km s$^{-1}$; this indeed reaches larger $\lx$
values. Finally, the bottom right panel shows the average of the
histograms of the three models, and shows how this average reproduces
the observed histogram reasonably well.  Overall, this analysis shows
that the large observed variation in $\lx$ at similar optical $\lb$
has another contributing factor, that is nuclear activity, in addition
to those already put forward.

Another useful diagram for an observational comparison is given by the
relationship between $\lbh$ and $\lx$; this is shown in Fig.~\ref{f3},
considering all the available temporal outputs, for model \bd. The
analogous figure for model \bt is very similar, with the only
difference being the broader temporal extension of the bursts.  One
can recognize the interburst times, when the galaxy is replenished by
stellar mass losses, as periods with $\lbh$ almost constant, and $\lx$
increasing (which produces almost vertical lines).  During outbursts,
the soft $\lx$ and $\lbh$ abruptly increase and then decrease, which
produces loops running clockwise on the right of each vertical line;
these loops are occupied for a very short time.  As time proceeds, the
vertical lines (the interburst periods) shift towards lower $\lbh$.
The final quiescent period is described by a line (the most crowded
with numbers) where $\lbh$ and $\lx$ both decrease, though $\lbh$
decreases faster than $\lx$.  The more the outbursts are confined at
earlier epochs, the more the final weak correlation between $\lx$ and
$\lbh$ extends with time; instead, the more the outbursts extend
towards the present epoch, the more a trend is expected for $\lx$ to
increase with $\lbh$ with a large scatter around it (as produced by
the vertical lines, where galaxies reside most of the time, and by the
loops).  For a sample of early type galaxies of the local universe,
the relationship between $\lx$ and the nuclear emission $L_{\rm
  X,nuc}$ indeed shows a weak trend, dominated by a large scatter
(Pellegrini 2010); in the present framework, such an observation could
be explained with many galaxies being still in the phases made of the
vertical rise followed by the loop.  The comparison cannot be pushed
farther, though, since the $\lbh$ values -- when converted to an X-ray
band -- are typically larger than the observed $L_{\rm X,nuc}$
(Sect.~\ref{nuc}).

\subsection{Temperature evolution and $\lx-\ta$}\label{ttt}

\begin{figure*}
\hskip -1truecm
\includegraphics[angle=0.,scale=0.55]{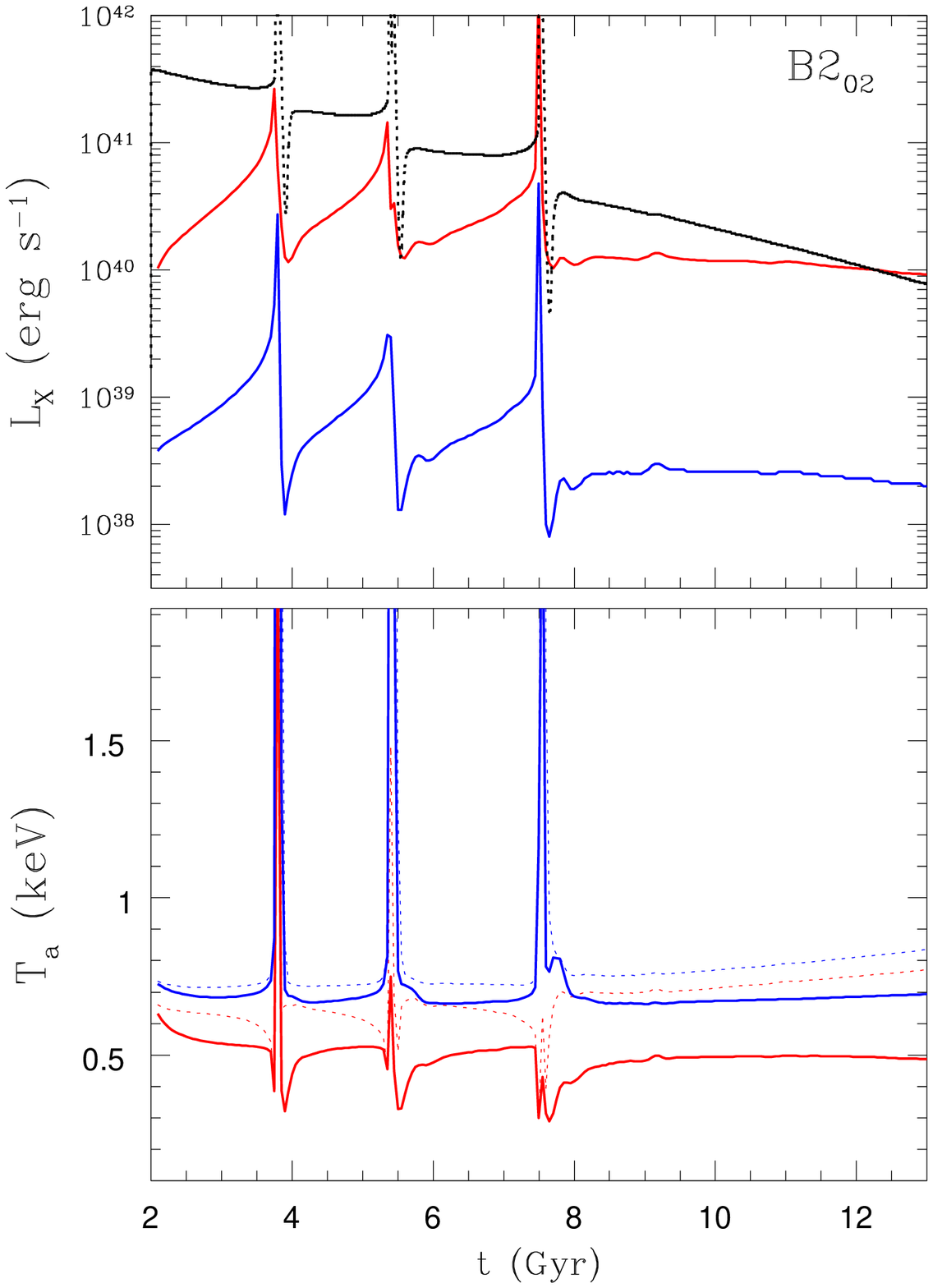}
\hskip -3truecm
\includegraphics[angle=0.,scale=0.55]{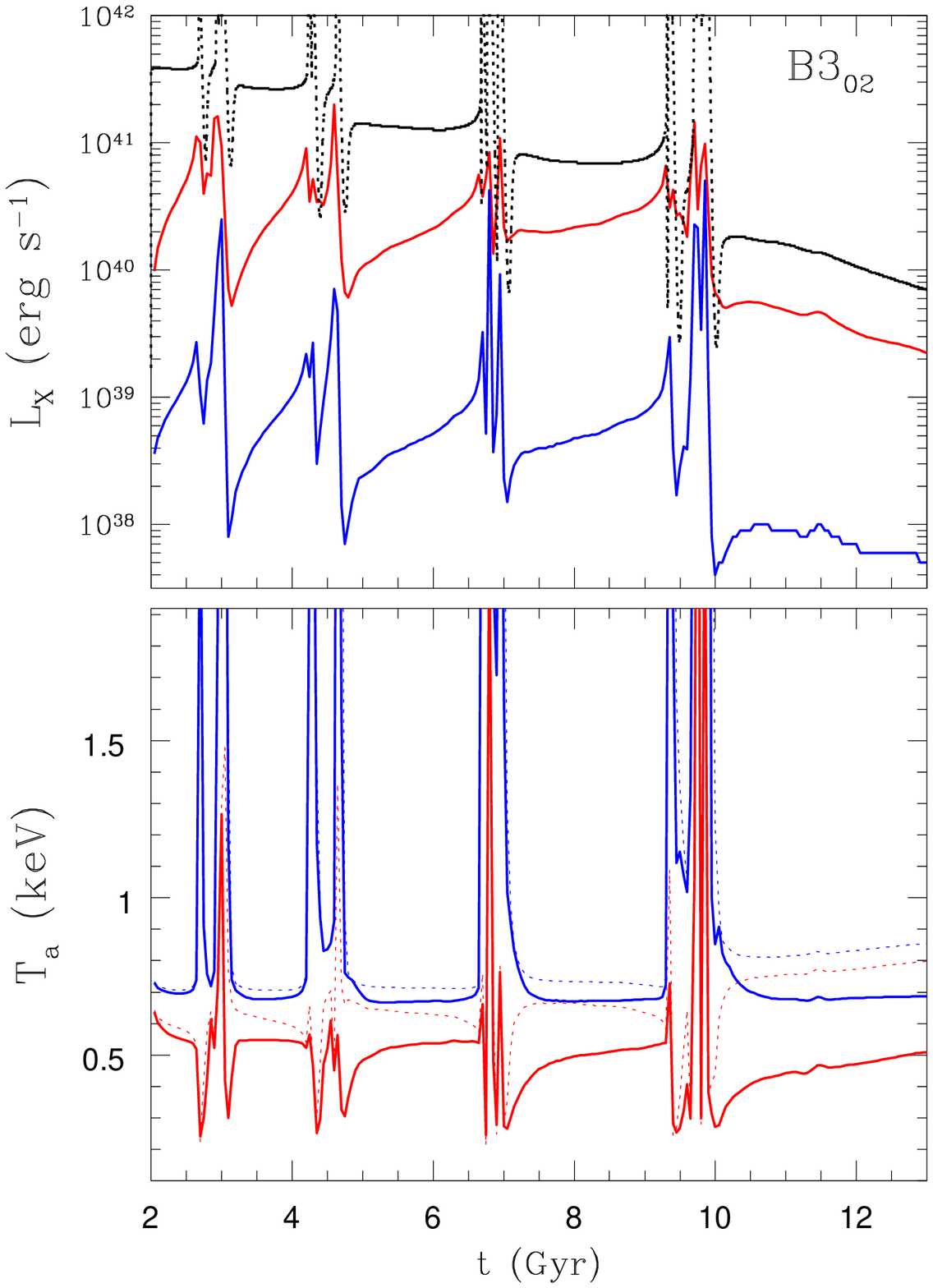}
\caption{Time evolution, shown with solid lines, of the ISM X-ray luminosity $\lx$ (upper
panels) and emission-weighted temperature $\ta$ (lower
panels), both calculated within an aperture of $10\re$,
for model \bd (left panels) and \bt (right panels). Red
and blue lines refer to the 0.3-2 keV and 2-8 keV bands. For
reference, the black dotted line in the upper panels shows $\lbh$
scaled down by a factor of 2000 from Fig.~\ref{f1}.  In the bottom
panels, the dotted lines show $\ta
(\re)$; in each band, $\ta(\re)$  is higher than $\ta(10\re)$.
Note the characteristic opposite trend of the red and blue temperatures
during the bursts, a clear sign of the coexistence of hot (central
bubble) and cold (radiative shells) ISM phases.
Temperatures
computed over the whole 0.3--8 keV energy interval (not
shown here) are always very close to those weighted with the 0.3--2 keV
emission, except during the burst times. See Sect.~\ref{global} for
more details.}
\label{f2}
\end{figure*}

\begin{figure}
%\vskip -6truecm
\hskip 2truecm
\includegraphics[angle=0.,scale=0.65]{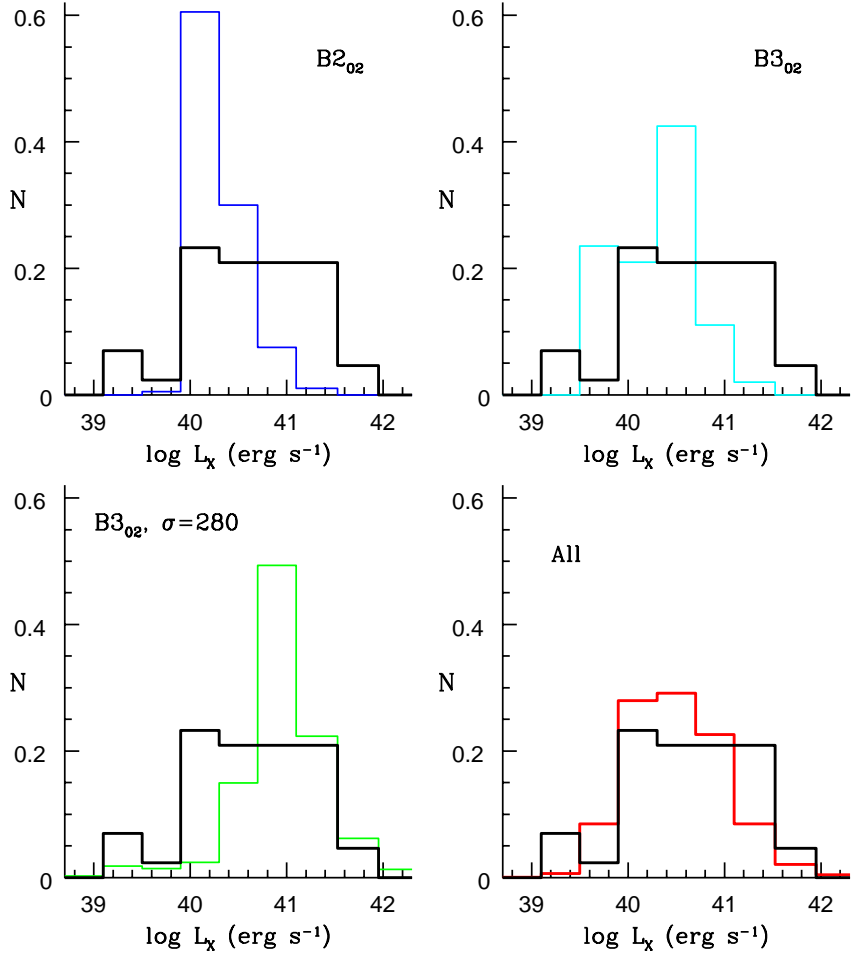}
\caption{Normalized histograms of the 0.3--2 keV gas emission within
 $10\re$ during the epoch from 2 to 12 Gyrs (colored lines), for the
 models indicated in each panel, and for their average (bottom
 right); the variant of \bt with $\sigma $ increased to 280 km
 s$^{-1}$ (bottom left panel) is taken from Ciotti \& Ostriker 2011.
 The histograms are compared with the histogram of observed
 luminosity values (converted to the 0.3--2 keV band), for non-AGN
 and non-central cluster or group members, with log$L_B$ in the range
 from 10.5 to 10.8 (black line; values from the $ROSAT$ sample of
 local early type galaxies, O'Sullivan et al. 2001; their estimate of
 the stellar source contribution has been subtracted to the observed
 $L_X$).}
\label{his} 
\end{figure}

\begin{figure}
\hskip -1 truecm
\includegraphics[height=26cm,width=22cm]{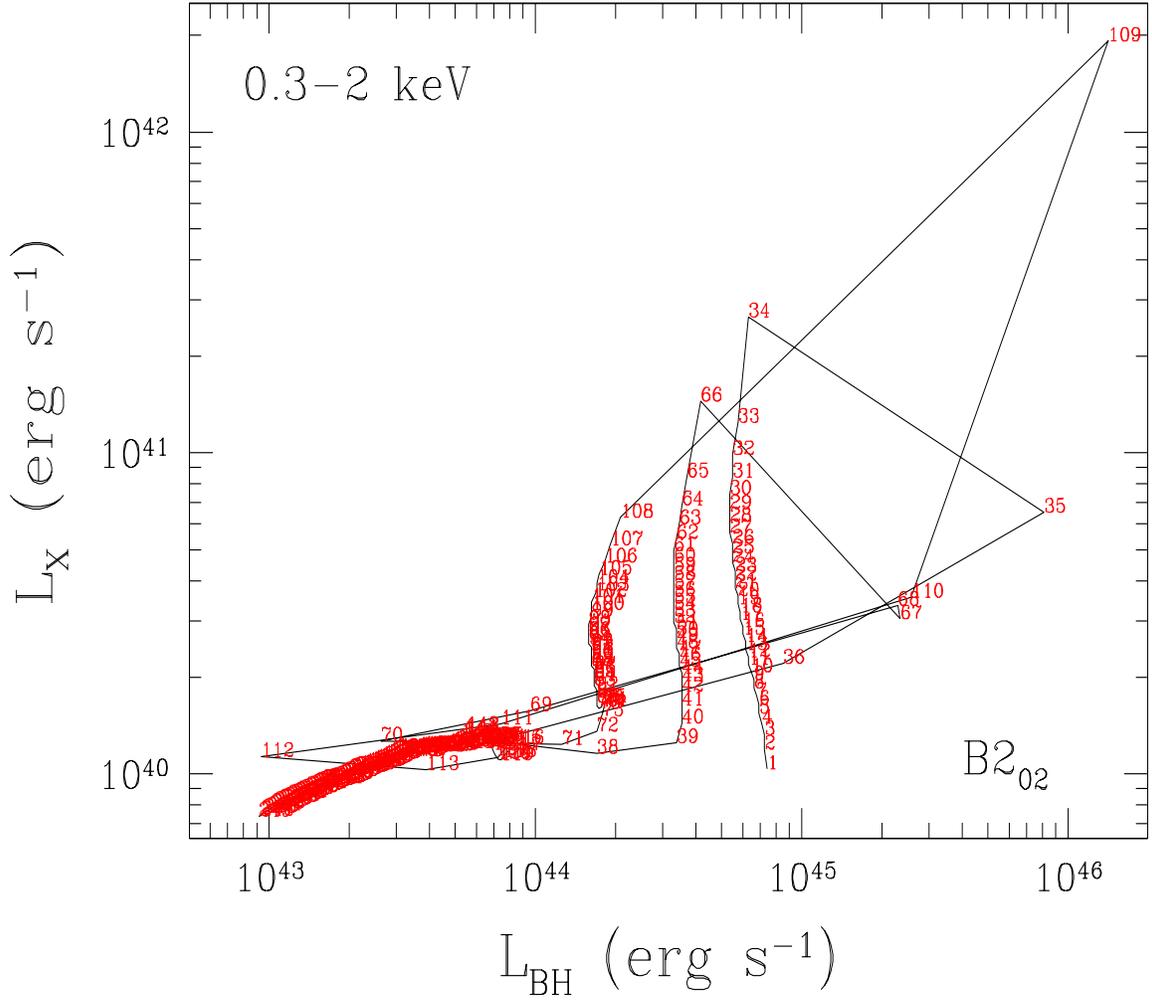}
\vskip -9.2truecm
\caption{The total 0.3-2 keV gas emission versus the bolometric
nuclear radiative output $\lbh$, for model \bd.  The time evolution
from 2 to 14 Gyrs corresponds to the increasing numbers along the
curve; the time sampling is uniform, i.e., every two subsequent points
are  50 Myr distant in time.}
\label{f3}
\end{figure}

Another important global property of the ISM typically observed is the
emission weighted aperture temperature $\ta$ (eq.~\ref{ta}), here 
calculated within the optical $\re$, and within $10\re$.  The
time evolution of these temperature diagnostics is shown in the bottom
panels of Fig.~\ref{f2}, in parallel with the luminosity evolution;
red and blue lines again refer to the 0.3--2 keV and 2--8 keV bands,
respectively.  Note that $\ta(10\re)$ in the two bands is
indistinguishable from the corresponding global emission weighted
temperature $<\tx>$ (that is calculated but not shown in
Fig.~\ref{f2}), as expected given that the density profile is steeply
decreasing outward (Paper III); thus $\ta(10\re)$ is a good proxy for
$<\tx>$. Also, temperatures computed for the whole 0.3--8 keV band
(not shown) are always very close to those weighted with the 0.3--2
keV emission, except for those short burst times during which there is
a very hot gas component. As for $\lx$, also for $\ta$ the temperature
peaks of \bt are significantly more structured in time than those of \bd.

Three main characteristic features of $\ta$, common to this class of
models, can be pointed out.  The first is the complex behavior of
$\ta$ during outbursts, with variations going in opposite
directions for the two bands. This is due to the coexistence of hot
(the central bubble) and cold (the radiative shells) ISM phases during
the bursts, as revealed by the temporal evolution of the radial
profiles of the gas density and temperature (Paper III). The sharp and
high peaks in the hard band (blue) correspond to the onset of very hot
regions at the center, while the decrements in the soft band (red) are
due to a dense cold shell created immediately before the major burst,
and to cold gas accumulated by the passage of radiative shock waves
produced by the outburst (see also Sect.~\ref{tempx}).

A second, observationally relevant feature is that in each band
$\ta(\re)$ is higher than $\ta(10\re)$, which is especially evident in
the interburst periods (Fig.~\ref{f2}). For example, in the quiescent
phases, both models show a similar $\ta(10\re)\simeq 0.4-0.5$ keV in
the soft band, while $\ta(\re)\sim 0.7$ keV. This is explained by the
radial temperature distribution decreasing outward
(Sect.~\ref{tempx}).

\begin{figure}
\hskip -1 truecm
\includegraphics[height=26cm,width=22cm]{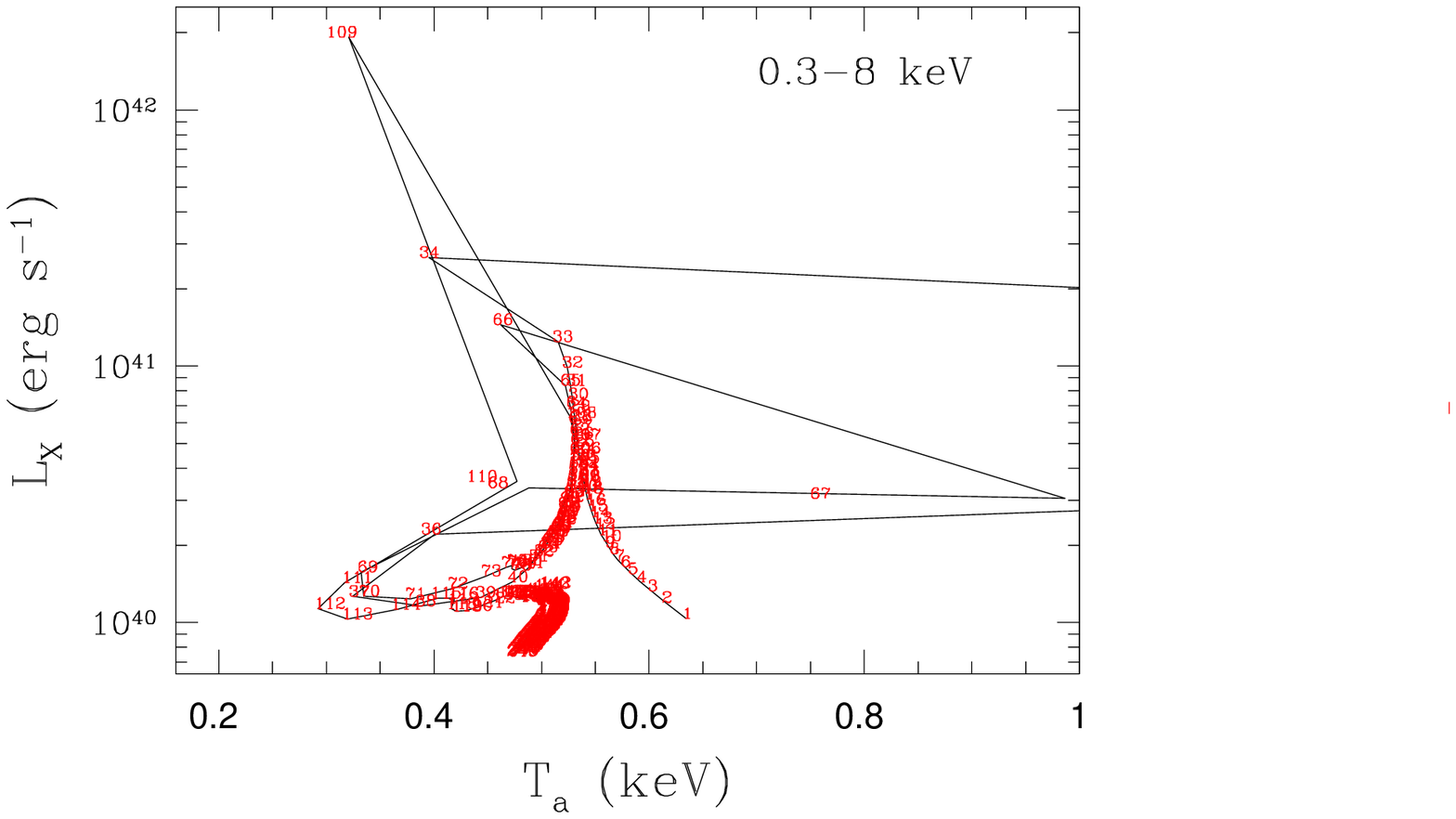}
\vskip -9.2truecm
\caption{The total gas emission versus the aperture
temperature (within $10\re$), calculated for the 0.3-8 keV band, for model \bd.
%(left panel), and the total 0.3-8 keV band (right panel)
As for Fig.~\ref{f3}, the time evolution
from 2 to 14 Gyrs corresponds to the increasing numbers along the
curve; the time sampling is uniform, with time steps of 50 Myr.}
\label{uff}
\end{figure}

Finally, for model \bd Fig.~\ref{uff} shows the relationship between
$\lx$ and $\ta(10\re)$, both calculated for the 0.3--8 keV band; such a
diagram is often produced in observational works, for galaxies of all
$\sigma $ (e.g., Pellegrini 2011).
During the long interburst epochs, $\lx$ increases with
little variation of $\ta$. In the short burst episodes, $\lx$ and
$\ta$ reach a maximum, and then follow a clockwise loop on the right,
reach back the original $\ta$ value, move left of it, and follow a
counterclockwise smaller loop. Then the cycle repeats with $\lx$
increasing and $\ta$ almost constant. During the last few Gyr, $\lx$ remains
at $\sim 10^{40}$ erg s$^{-1}$ and $\ta \sim 0.5 $ keV.

We now compare these results with observations. In the largest sample
of global, soft X-ray emission weighted temperatures, derived from
$ROSAT$ observations, $<\tx>$ is in the range $\sim 0.4-0.8$ keV, for
galaxies with $\sigma\simeq 260$ km s$^{-1}$, as for the model galaxy
(O'Sullivan et al.~2003); $\ta(10\re)$ of the models during
quiescence, in the soft band, is within this range, though on
its lower end.  However various factors tend to bias-high these
observed temperatures, as incomplete subtraction of the hard stellar
emission due to binaries, or hard AGN emission; in addition, many of
the sample galaxies reside in high density environments, with possible
contamination from the hotter group/cluster medium, and a temperature
profile that is commonly rising outwards (e.g., Diehl \& Statler 2008,
Nagino \& Matsushita 2009), a behavior opposite to that of the models,
that refer to isolated galaxies (Sect.~\ref{tempx}).

\begin{figure*}
\vskip -5truecm
\hskip -1truecm
\includegraphics[angle=0.,scale=0.6]{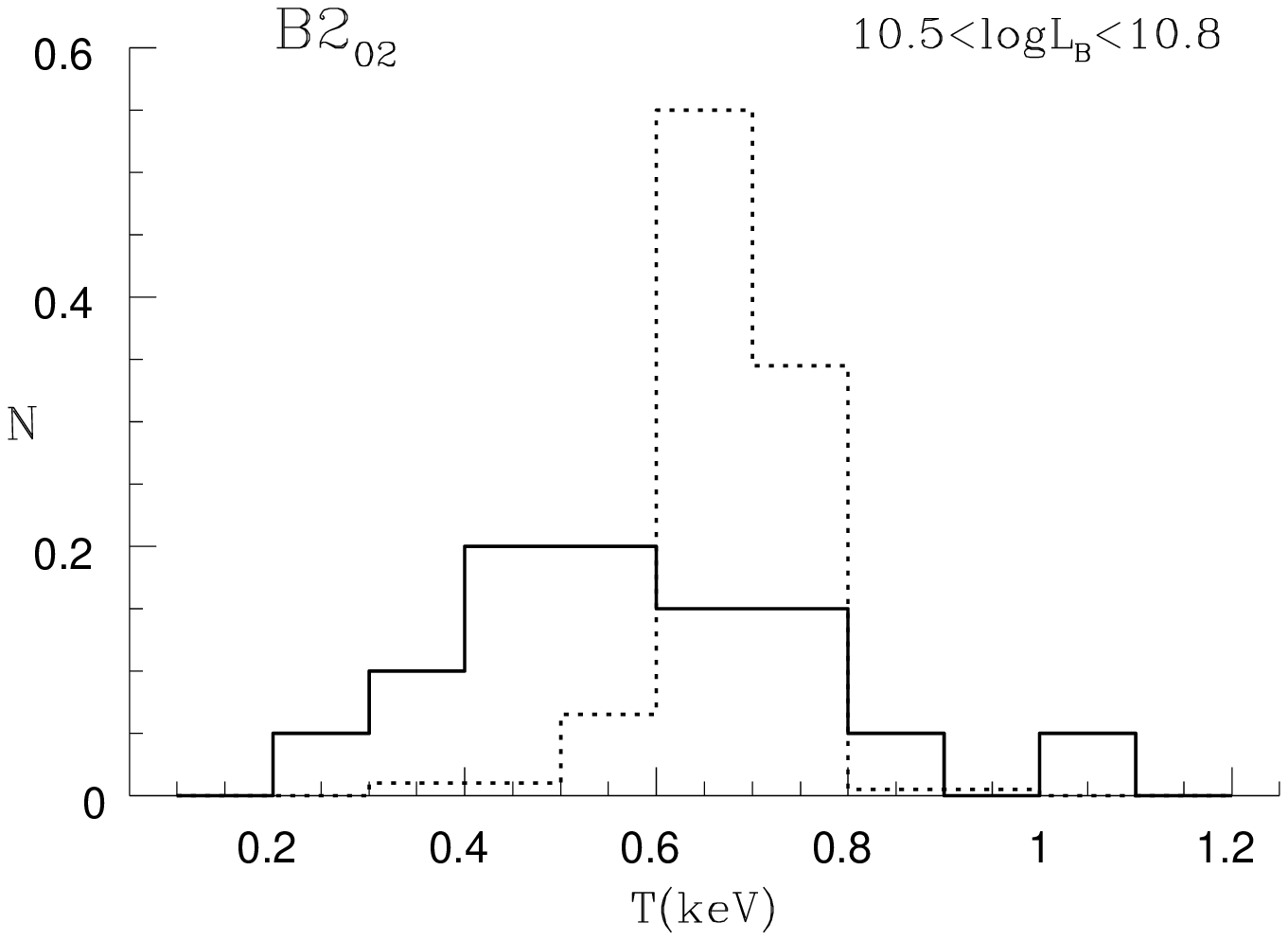}
\hskip -3.5truecm
\includegraphics[angle=0.,scale=0.6]{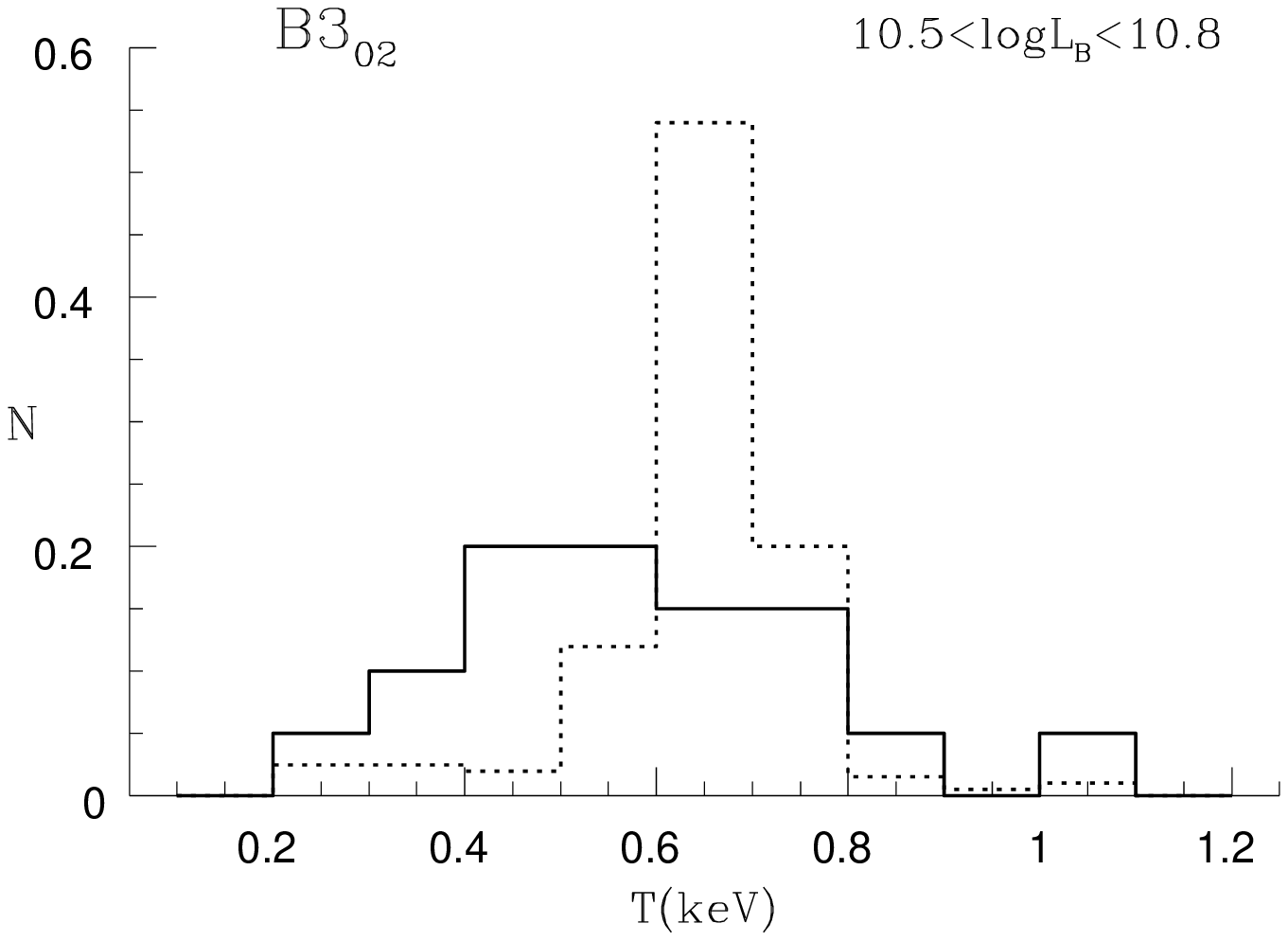}
\caption{Normalized histogram (dotted lines) of the 0.3--8 keV emission weighted
temperature $\ta(\re)$, during the epoch from 2 to 12 Gyrs, for the model indicated
in each panel; the histograms are calculated as for Fig.~\ref{his}.
The histograms of the models are compared with that derived 
for $\ta(\re)$ from $Chandra$ data (Athey 2007), for a subsample of 20 ellipticals
in the log$L_B$ range from 10.5 to 10.8 (solid line).
See Sect.~\ref{ttt}.}
\label{hisT} 
\end{figure*}

$Chandra$ observations, with large sensitivity and much higher angular
resolution ($\sim 1^{\prime\prime}$), allowed for an accurate subtraction of all the AGN and
stellar sources contributions from the total emission, giving
measurements of the gas temperature of unprecedented accuracy (e.g.,
Boroson et al. 2011). For example, Boroson et al. derived global,
0.3--8 keV emission weighted temperatures, for a few galaxies with
$\sigma \sim 260$ km s$^{-1}$, ranging from 0.3 to 0.6 keV;
$\ta(10\re)$ of the models agrees very well with this result, falling
in the middle of the observed range. Coming to temperature estimates
for more central regions, Athey (2007) derived 0.35--8 keV emission
weighted temperatures within $\re$, for 53 galaxies with $Chandra $
observations. For a selection of 20 galaxies with $\lb$ similar to
that of the model galaxy, from log $\lb (L_{B,\odot})=10.5$ to 10.8,
the average temperature is $0.61\pm 0.03$ keV (calculated weighting
each measurement with its uncertainty). Figure~\ref{hisT} shows a full
comparison between the distribution of these temperatures and
$\ta(\re)$, calculated for the 0.3--8 keV band, during the evolution
of \bd and \bt.  The model $\ta(\re)$
tends to be concentrated in the upper half of the observed
distribution; this result, if confirmed with a larger set of 
simulated galaxies, indicates that heating in the
central galactic region of the models may be too efficient.

In conclusion, the global temperatures of the models fall in the
middle of the range of values recently observed, while the
model $\ta(\re)$ reproduces easily the larger observed values
and less easily the lower ones.

\section{Projected quantities: temperature and brightness profiles}
\label{proj}

We consider here the radial profiles of the ISM temperature and surface
brightness, as they would be revealed for the models by observations.
For simplicity we restrict the discussion to model \bd, whose sharp
bursts allow for an easier presentation; the results are substantially
similar for \bt.  The profiles are constructed using
Eqs.~\ref{sig}--\ref{conv}, both during the quiescent phases, that
occupy most of the ISM evolution, and during an outburst.  The
recurrent burst phases are temporally limited, but represent a central
aspect of the models, thus they are devoted special attention.
Statistically, the feedback features should be present, and possibly
revealed by current X-ray observations, in $\simeq 5-10$\% of the
isolated galaxies with $\lb$ similar to that of the models (Sect.~\ref{brilx}
below). In the following we present snapshots of the projected
profiles during quiescence, at an age of 3, 6.5, 9 and 12 Gyr, and
centered on the last outburst at 7.5 Gyr.

\subsection{Temperature profiles}
\label{tempx}

\begin{figure*}
\vskip -6.truecm
\hskip -1. truecm
\includegraphics[height=14cm,width=12cm]{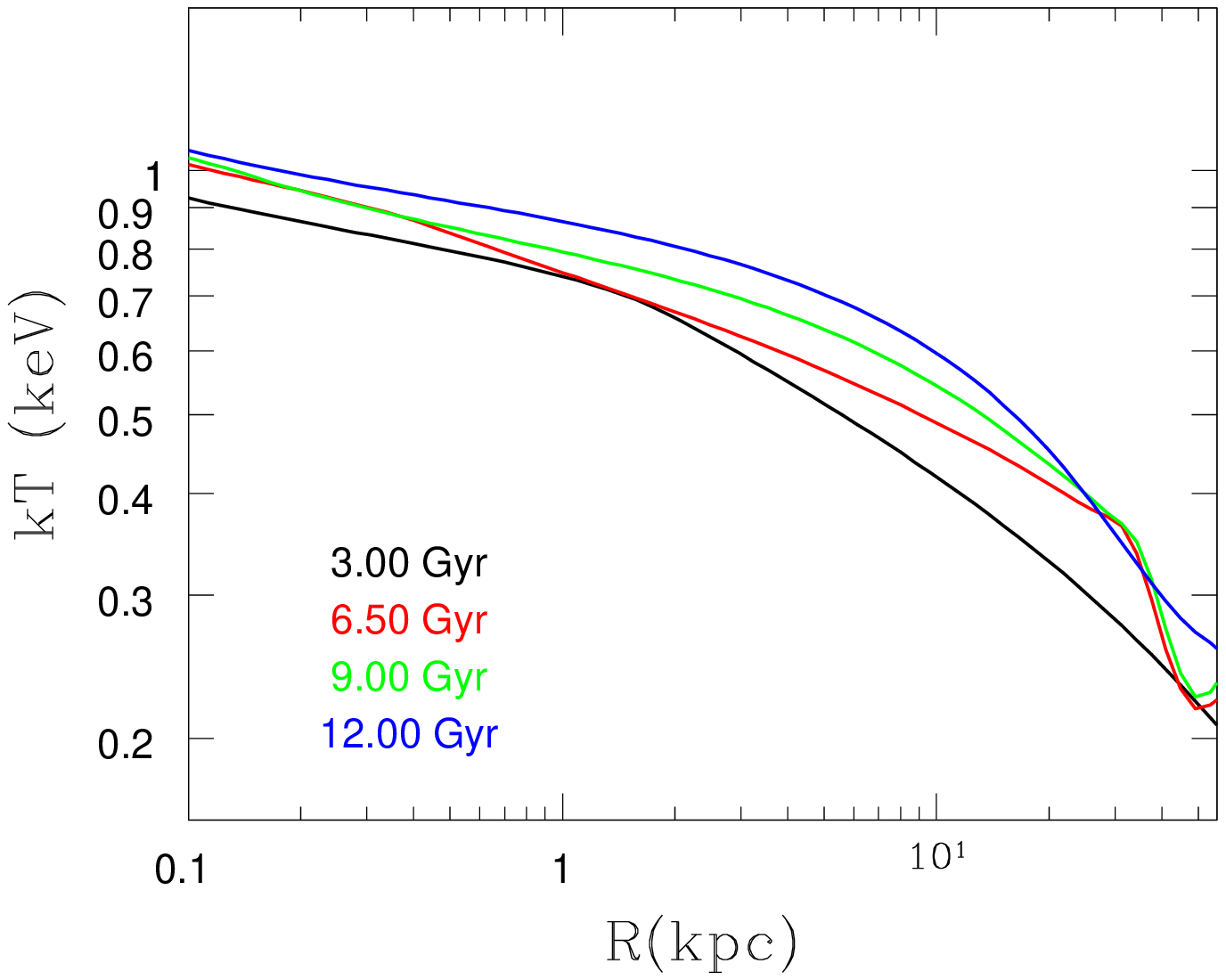}
%\vskip -5.5truecm
\hskip -3.5truecm
\includegraphics[height=14cm,width=14cm]{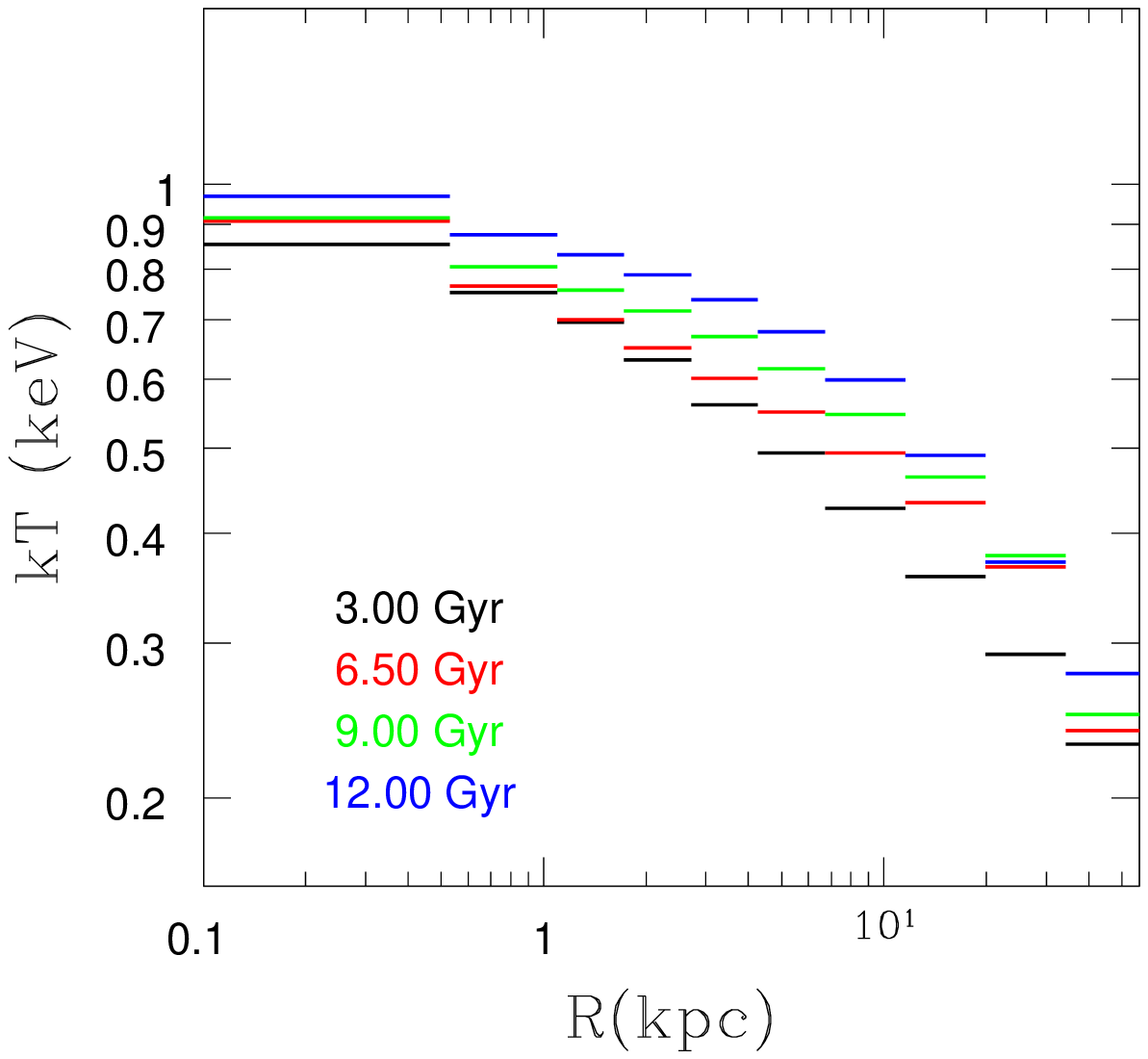}
\caption{ {\it Left panel}: radial profile of the 0.3--8 keV
emission-weighted projected temperature $\tp(R)$ (Eq.~\ref{tprof}),
during the interburst times indicated in the panel, for model \bd. The
temperature increases with time, due to the secular increase of
the ISM specific heating, and the growing MBH mass, which modifies
the local gravitational field and the central stellar velocity dispersion 
(see Sect.~\ref{tempx} for more details). The sharp drop
in the red and green profiles at $\approx 40$ kpc is due to
disturbances produced by the outbursts at 5.5 and 7.5 Gyr
(Fig.~\ref{f1}) that are still traveling outward.  {\it Right panel}:
the corresponding aperture temperature profiles $\ta (R)$, obtained
from averaging $\tp(R)$ using the surface brightness (Eq.~\ref{ta}).
The bin-width increases going outward in the galaxy, to reproduce the
best  observed profiles of nearby galaxies.}
\label{f5}
\end{figure*}

The emission weighted projected temperature profiles $\tp(R)$ during
quiescent interburst times are smooth, with the temperature
monotonically decreasing for increasing radius (Fig.~\ref{f5}).  From
an age of 6 Gyr onwards, the temperature keeps at $\simeq 1$ keV at a
radius of $\simeq 100$ pc, and at $\sim 0.4$ keV at a radius of
$\simeq 30$ kpc.  Figure~\ref{f5} (right panel) shows the
corresponding aperture temperature profiles $\ta (R)$ calculated in
bins  (i.e., in Eq.~\ref{ta}
the numerator and denominator are
evaluated over radial bins), chosen to reproduce
the typical binning used for $Chandra$ observations of nearby galaxies
(Humphrey et al.~2006, Diehl \& Statler 2007). 

Temperature profiles with negative gradients had already been found
for gas inflows in steep galactic potentials without feedback, due to
compressional heating (e.g., Pellegrini \& Ciotti 1998).  As typical
of models without a central MBH, in that case the temperature also had
a central drop.  In the present computations, instead, the combined
effects of the gravitational field of the MBH, and of the high
injection temperature of the stellar mass losses (a consequence of the
stellar velocity dispersion that is enhanced by the presence of the
MBH, within its sphere of influence), keep the gas temperature
increasing approaching the center, even outside the burst episodes
(see also Pellegrini 2011).  A temperature profile that keeps
smoothly decreasing towards the
center for a long time, as expected in classical cooling flows, is never shown by 
the model runs.

With time increasing, the value of the central temperature does not
increase significantly after 6 Gyr, since it is mainly determined by
the gravitational effects of the MBH, whose mass remains approximately
constant.  The external $\tp(R)$ values instead steadily increase with time,
since they are determined by the SNIa's heating; the latter has a secular
trend that produces a time-increasing specific heating of the mass return
rate (i.e.,  $L_{\rm SNIa}/\dot M_*\propto t^{0.2}$, where $L_{\rm SNIa}$ is the
energy injected per unit time by SNIa's supernova explosions, and
$\dot M_*$ is the stellar mass loss injected per unit time; e.g., see
Pellegrini 2011).

We now focus on the evolution during the last outburst
(Fig.~\ref{f6}).  As typical, starting from the unperturbed profile, a
shell of denser gas creates, in this case at a radius of $\simeq 800$
pc, particularly evident as a dip in the otherwise monotonically
decreasing profile (see the -2 Myr red line in the top left
panel). The shell falls towards the center, progressively cooling, so
that the soft X-ray emission weighted temperature decreases
(Fig.~\ref{f2}, bottom panels).  A first AGN burst is produced before
the shell reaches the center, and the resulting outward moving shock
empties and heats the gas in the inner regions of the galaxy, on a
time scale of $\approx 1$ Myr.  The subsequent snapshot in
Fig.~\ref{f6}, at +6 Myr after the first burst, shows the high central
temperature typical of the outburst phase: $\tp(R)\sim $few keV within a
radius of $\sim 50-100$ pc.

When the shock enters the radiative snow-plow phase, the associated
secondary dense shell interacts with the first one still falling, and
a series of direct and reflected shock waves are produced, accompanied
by accretions events and star formation.  The profiles
between -2 Myr and +18 Myr (not shown) are very irregular, and
consisting of a series of dips propagating outwards, with the
temperature in the central region quickly and alternately increasing
and decreasing. As already noticed for Fig.~\ref{f2}, both hotter and
colder regions are continuously created, the hottest ones within $\sim
1$ kpc (mainly due to shocks), and the coldest ones due to the cooling
gas in the shells.

Each sub-burst is stronger than the previous one; finally, a last
shock from the center concludes the phenomenon, halting star formation
and leaving behind a very hot nucleus (the +66 Myr red line), while
emptying the rest of the galaxy [see also the $\Sigma (R)$ profile at
+66 Myr in Fig.~\ref{f11}]. Then, during the following $\simeq 100$
Myr, the gas resumes the temperature typical of quiescent times (the
+202 Myr green line).  The inner very hot phase lasts for $\lsim 0.1$
Gyr.

During the bursts, cosmic rays are shock-accelerated, and the inner regions look
similar to a gigantic supernova remnant.  Only thermal X-rays are
considered  here, while those due to synchrotron emission from
accelerated particles due to shocks were not
computed (see Jiang et al.~2010).

The profiles in the right panels of Fig.~\ref{f6} show the binned
aperture temperature profiles, for the same times as for the left
panels. The binning smears the small-scale features, but the major
distinctive characteristics, as the temperature dip when the first
shell forms, and the large temperature drop outside the center, at +18
Myr and +66 Myr, are still well detectable.

\subsubsection{Comparison with observed profiles}\label{tcomp}

\begin{figure*}
\vskip -4.7truecm
%\hskip -0.7truecm
\hskip -0.9truecm
\includegraphics[height=14cm,width=12cm]{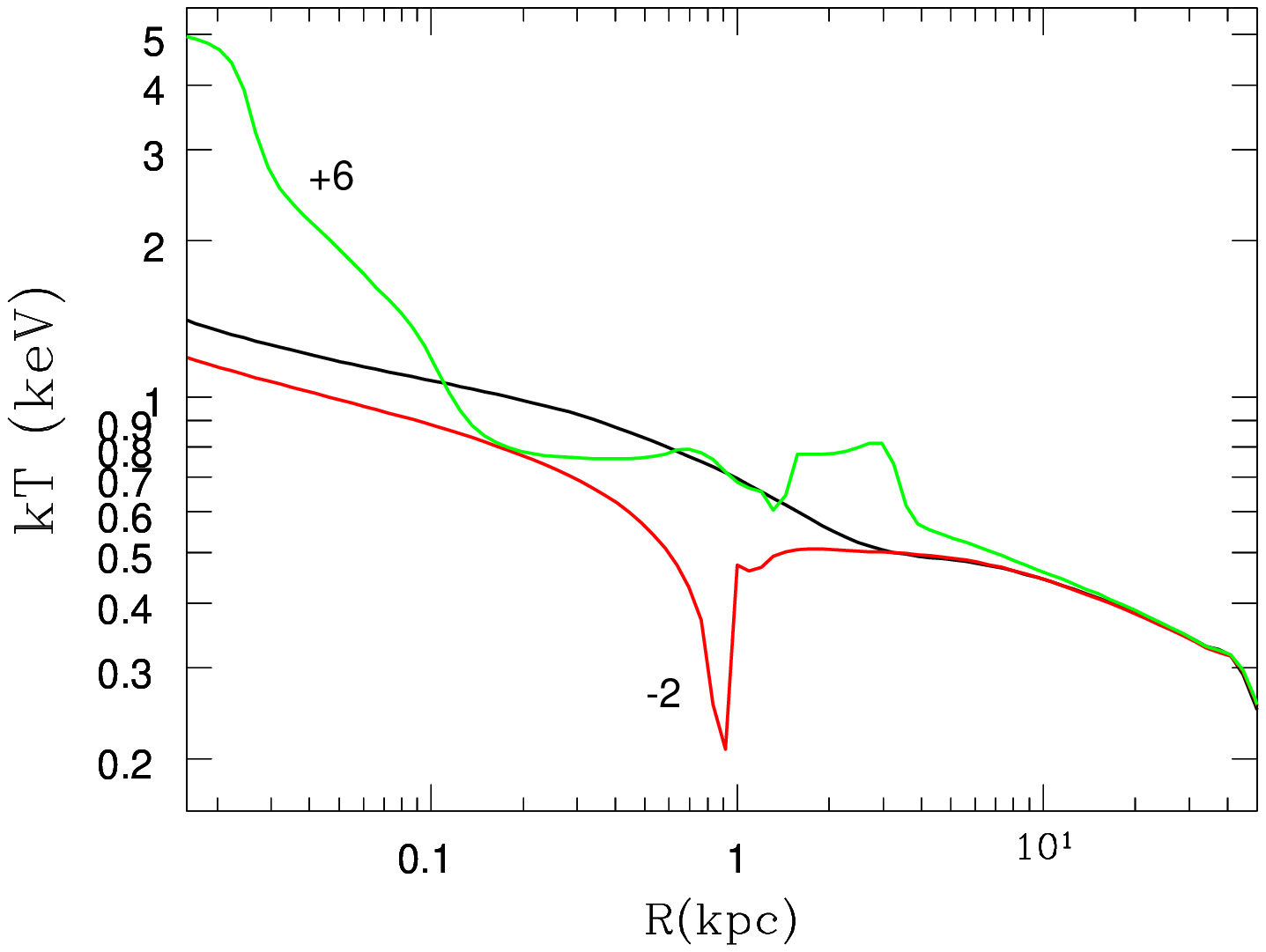}
%\vskip -0.1truecm
\hskip -3.5truecm
\includegraphics[height=14cm,width=14.5cm]{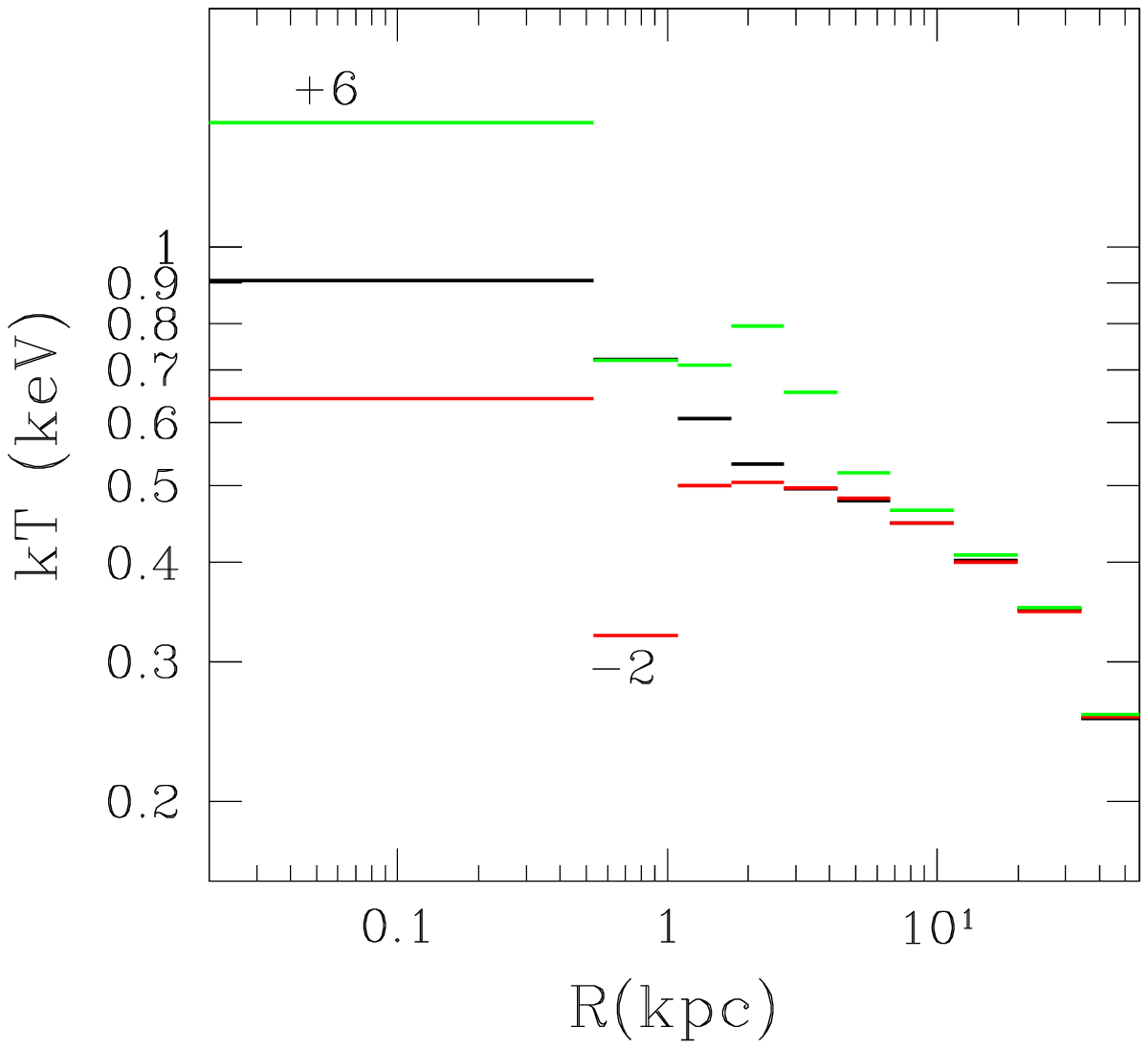}
%\vskip -5.8truecm
\vskip -4.8truecm
\hskip -0.9truecm
\includegraphics[height=14cm,width=12cm]{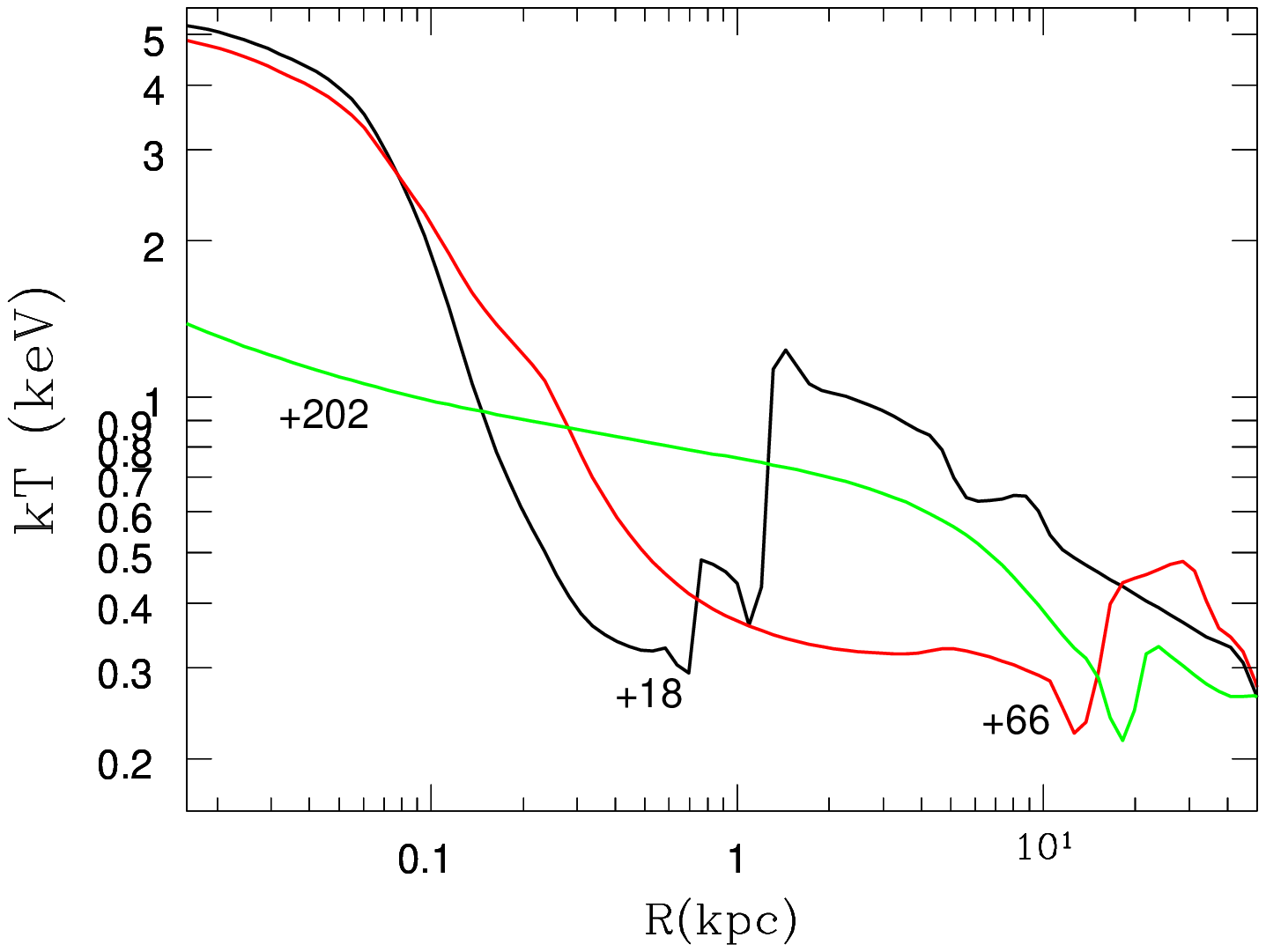}
\hskip -3.6truecm
%\vskip -0.1truecm
\includegraphics[height=14cm,width=14.5cm]{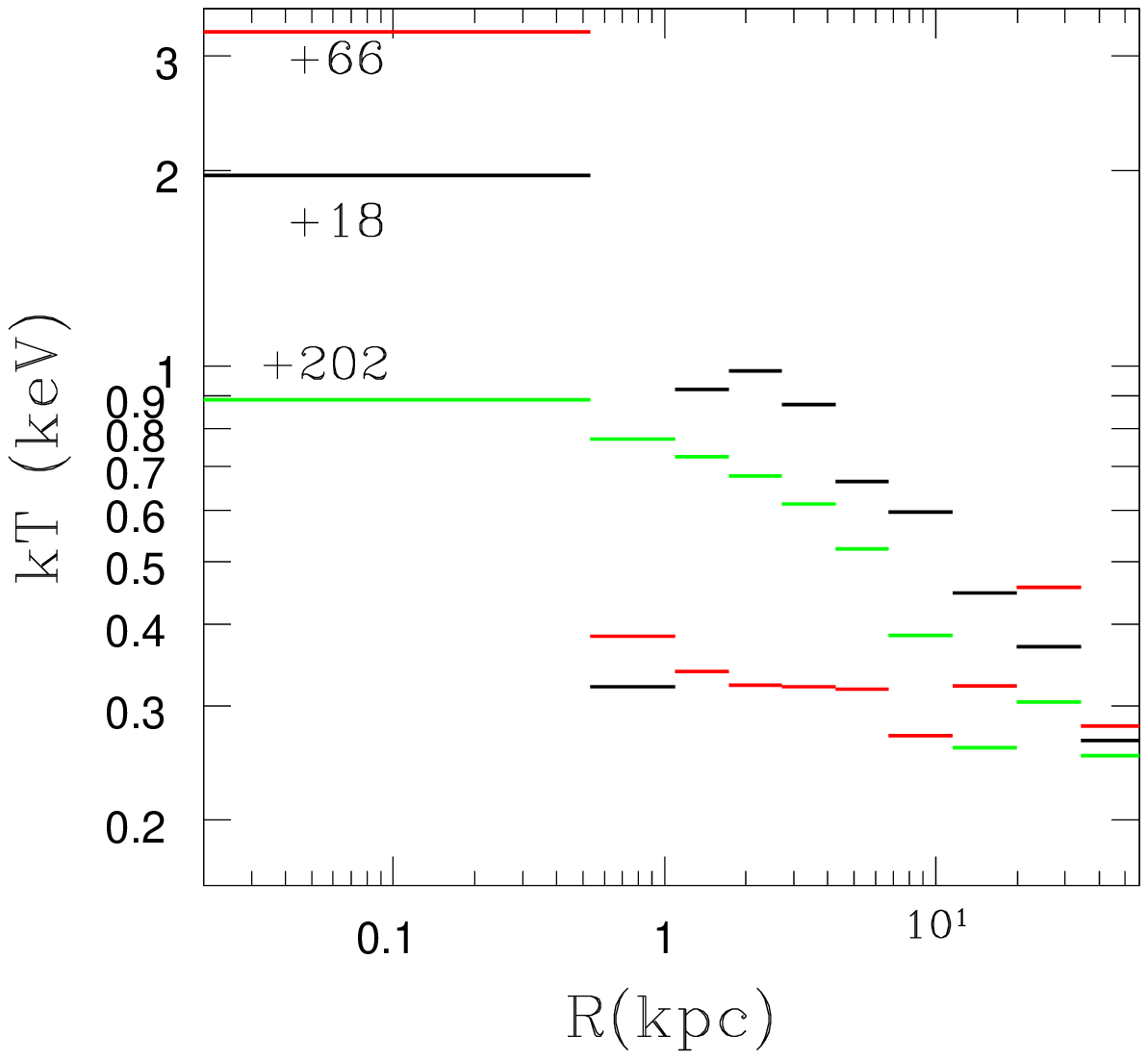}
%\vskip -2truecm
\caption{{\it Left panels}: radial profiles of the emission-weighted
projected temperature $\tp(R)$ in the 0.3-8 keV band, during the
last major burst of model \bd (at $\simeq 7.498$ Gyr).  The numbers
near the lines indicate the times (in Myr) calculated with respect
to the outburst; in the top panel the black line shows the
unperturbed profile before the outburst (at a time of 7.400 Gyr). {\it Right panels}: the corresponding aperture
temperature profiles $\ta (R)$, averaged with the surface brightness
in bins with the same radial range adopted for Fig.~\ref{f5}.}
\label{f6} 
%tempbur(_6b).m in /newb202/burhi/B2... legge 200, 223 e 225 (panel a)
% e 228, 240 e 274 (panel b)
\end{figure*}

Negative radial gradients, as shown by the model temperature during
quiescence (Fig.~\ref{f5}), are common among ellipticals, as revealed
most recently by $Chandra$ observations (e.g., Kim \& Fabbiano 2003,
Humphrey et al.~2006, Sansom et al.~2006, Fukazawa et al. ~2006, Diehl
\& Statler 2008b, Nagino \& Matsushita 2009).  In a large collection
of temperature profiles (Diehl \& Statler 2008b, Athey 2007), those
cases with negative gradients show a temperature that roughly halves
from the innermost bin (that in general extends out to a radius of
0.5-a few kpc from the center) to the outer galactic region; this is
roughly the same behavior shown by the models.  Observed temperatures
range between 0.5 and 1 keV at the innermost radial bin, where the
model temperature is 0.8-1 keV (Fig.~\ref{f6}, right panels).
Interestingly, the galaxies with a negative gradient reside in the
field or in less dense environments, as the outer Virgo regions
(Matsushita 2001, Fukazawa et al.~2006, Diehl \& Statler 2008b), and
all have\footnote{Except for NGC6482, the remnant of a fossil group.}
$\lx<$few$\times 10^{40}$ erg s$^{-1}$, both characteristics that
match those of the models.

More complex temperature profiles are also common, and could
correspond to phases of AGN activity. For example, the so-called
``hybrid'' profiles show a central negative gradient, until the
temperature reaches a minimum, and an outer positive gradient; for
example in NGC1316, NGC4552, NGC7618 the temperature has a minimum of
$\sim 0.4$-0.5 keV at $\sim $ few kpc, and rises both going towards
the center (of $\sim 0.2$-0.3 keV), and going outwards (Diehl \&
Statler 2008b). Also in the models, after each major burst, when the
last shock is moving outwards and fading, leaving a hot, rapidly
cooling core, there is a drop in the temperature profile at a radius
of $\sim 1$ kpc or more (Fig.~\ref{f6}, bottom panels, black and red
profiles).  However, in the models the temperature reaches $> 1$ keV
at the center, larger than the observed values (the comparison also
depends on the binning used for the profiles, though). Interestingly,
preliminary results from 2D simulations (Novak et al. 2010), show
lower central temperatures than in Fig.~\ref{f6}, during outbursts,
due to the fragmentation of the cold shell, that causes a lower gas
compression while falling to the center.

Other observed profiles keep roughly isothermal, or are roughly flat
out to $\sim\re$ and then increase outward (e.g., Humphrey et
al.~2006, Diehl \& Statler 2008b, Nagino \& Matsushita 2009). Such a
positive outer gradient is shown by galaxies in high-density
environments, suggesting the influence of circumgalactic hot gas;
also, these galaxies have typically a larger hot gas content than the
models.  Environmental confining effects, currently not included in
the models, are expected to increase the gas luminosity, and produce
outer postitive gradients (see, e.g., Sarazin \& White 1987, Vedder et
al.~1988).

It has been proposed that galaxies could behave according to a
scenario where weak radio AGN distribute their heat
locally and host negative inner temperature gradients, whereas more
luminous radio AGNs heat the gas more globally through a jet or rising
bubbles, and produce a flat profile, or a positive gradient (Diehl
\& Statler 2008b). Our models during quiescence could correspond to the
weak AGN phase; also, after an outburst, the temperature profile can
be flattish (excluding the innermost bin) out to a few kpc, and 
show a positive gradient outer of this radius  (see the red line of Fig.~\ref{f6},
bottom right panel).  However, without a confining environment, an
exploratory investigation conducted by us 
shows that the kinetic heating of a bright radio
AGN could cause a major degassing, rather than just a reversal
of the temperature gradient from negative to flat or positive.  Likely, a gas
rich environment providing confinement for the galactic coronae is a
necessary condition to observe a bright (extended) radio source, and
the confinement in turn also helps produce a positive temperature
gradient in the outer galactic regions.  Both aspects (the jet and a
dense environment) will be implemented in the models in the future.

In conclusion, the addition of a jet to the simulations could have 
positive effects, if it heats the gas outside $\sim\re$,
and reduces the accretion rate and then $\lbh$ during
the stationary hot accretion phase (Sect.~\ref{nuc}); it should not
increase the temperature at the center, though, since this is already
on the upper end of those observed even during quiescence.

\subsection{Brightness profiles}\label{brilx}

\begin{figure}
\hskip -0.3truecm
\includegraphics[height=12cm,width=12.7cm]{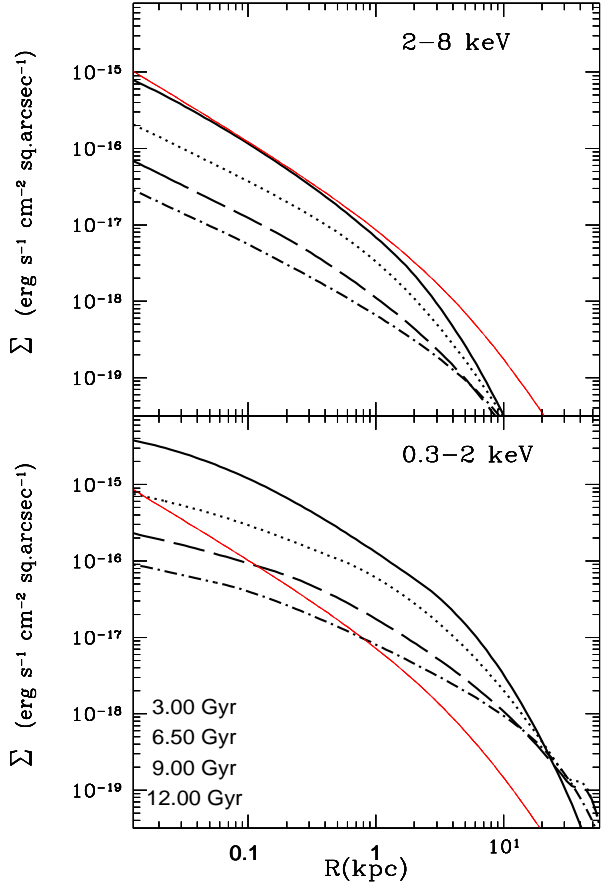}
\caption{The X-ray surface brightness profiles of the hot gas for
model \bd at quiescence, at the same times as for Fig.~\ref{f5} (given
in the bottom panel), for the hard band
(upper panel) and the soft band (lower panel).  The solid, dotted,
dashed and dot-dashed lines correspond to  increasing times.
The red line follows the optical profile and shows the fiducial unresolved
stellar emission due to low mass X-ray binaries, as would result
from a long observation of a galaxy of the same luminosity $\lb$
of the model (the line is normalized to give 20\% of the total
collective luminosity of these binaries, following the results for
local ellipticals observed with
$Chandra$, excluding the very hot gas rich ones, Boroson et al. 2011).}
\label{f9}
\end{figure}

Figure~\ref{f9} shows the evolution of the X-ray surface brightness
profile of the gas $\Sigma (R)$ for \bd, at the same
quiescent times of the temperature profiles in Fig.~\ref{f5},
for the soft (0.3-2 keV) and hard (2-8 keV) bands. In the hard band,
$\Sigma (R)$ is always comparable to or lower than a profile
representing the unresolved stellar emission due to low mass X-ray
binaries, calculated for a deep ($\gsim 200$ ksec) $Chandra$ pointing
of a galaxy of the same optical luminosity as \bd and distant $\lsim
20$ Mpc. Thus, hard emission during quiescent times would be difficult
to distinguish from the contribution of unresolved binaries, even with
$Chandra$. In both bands $\Sigma (R)$ becomes flatter in shape with
time increasing, since the emission level decreases mostly in the
central galactic regions (within $\re$).  This important effect is produced
by the nuclear outbursts, that remove gas from the center.

\begin{figure*}
%\vskip -0.8truecm
\hskip -1.5truecm
\includegraphics[scale=0.7]{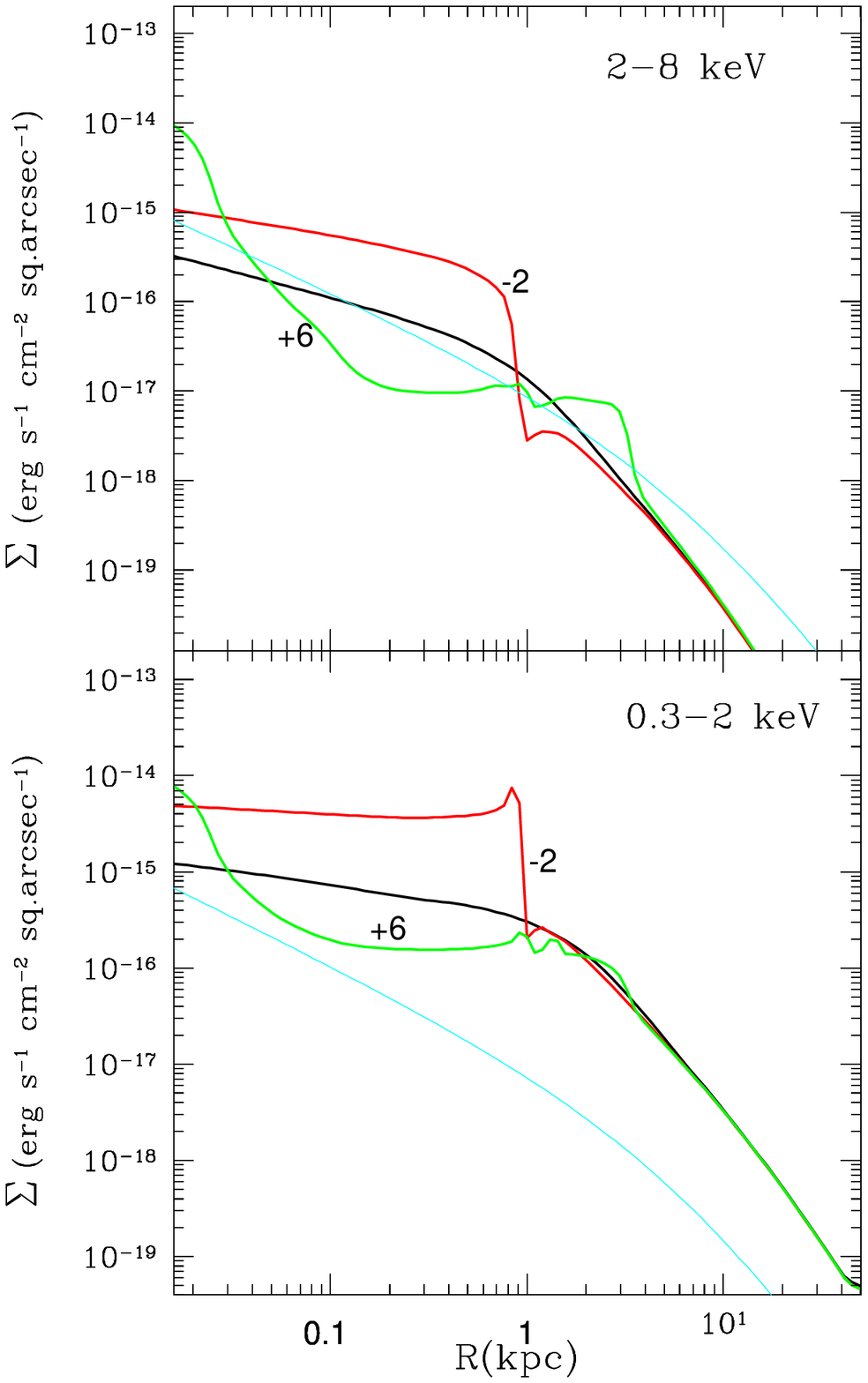}
\vskip -14.2truecm
\hskip 8truecm
\includegraphics[scale=0.7]{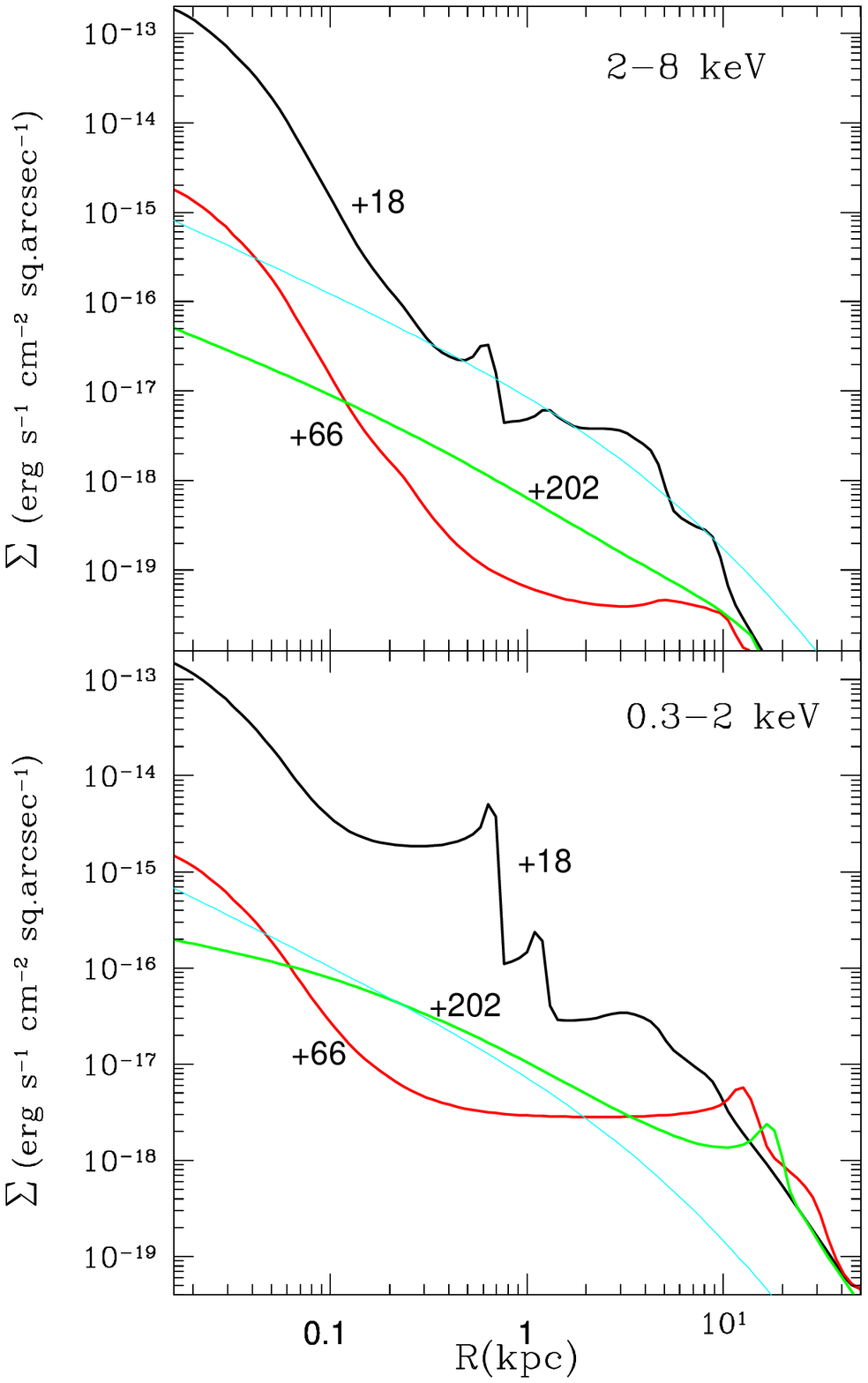}
\caption{The X-ray brightness profiles in the hard (top panels) and
soft (bottom panels) bands, during the last outburst of \bd
(occurring at 7.498 Gyr), for the same times as in Fig.~\ref{f6},
indicated in Myr close to each curve.  
The cyan line is an estimate for the unresolved binaries contribution
in the two bands, calculated as for Fig.~\ref{f9}. 
At -2 Myr the outburst is preparing and the shell is developing and
approaching the center; after the first outburst  the center hosts a
hot region (+6 Myr). A hot and dense central  region is present also
at +18 Myr, while an outward moving shock is still barely visible at +66 Myr and +202 Myr. }
\label{f11} 
\end{figure*}

\begin{figure*}
%\vskip -0.5truecm
\hskip -1.7truecm
\includegraphics[scale=0.7]{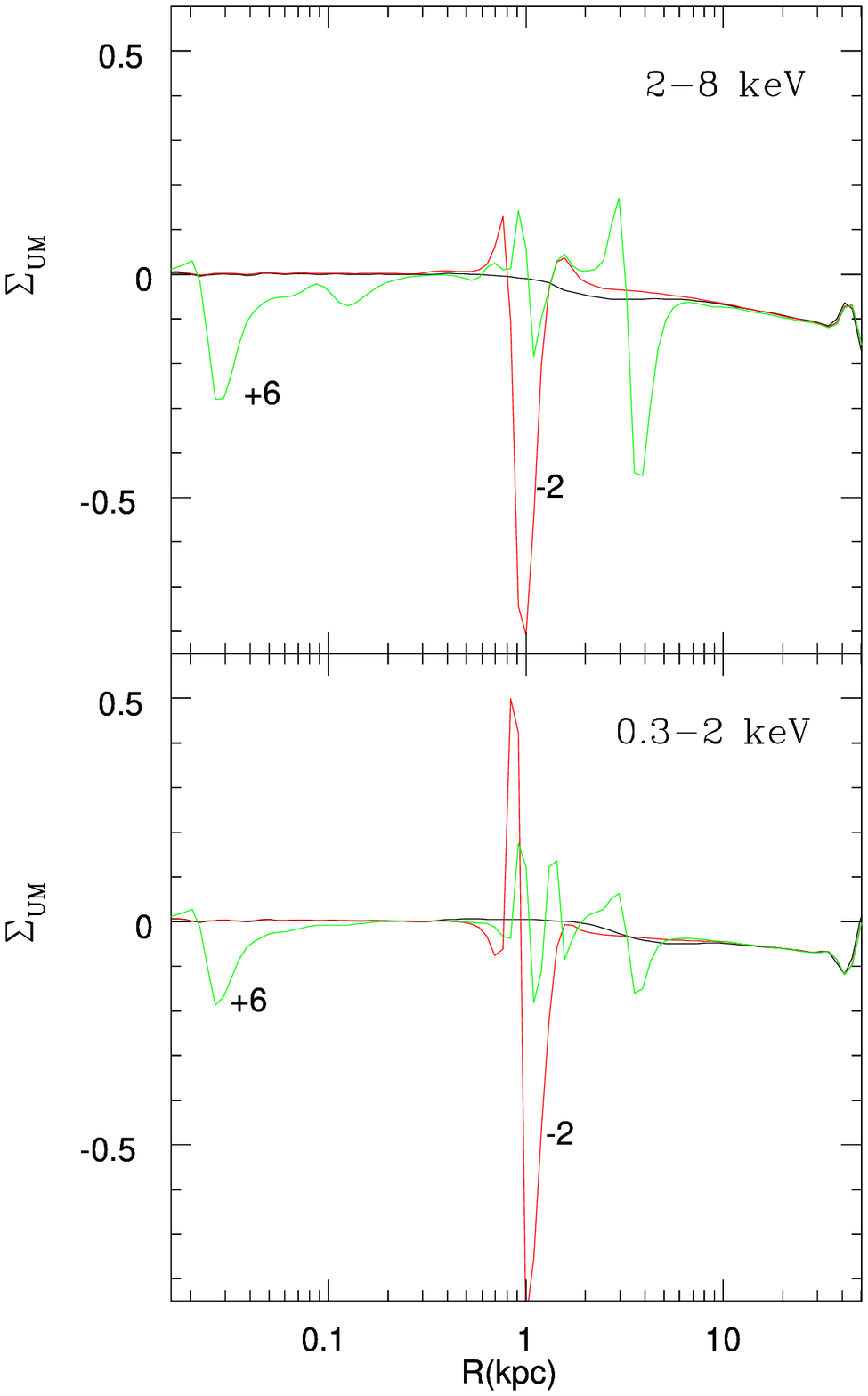}
\vskip -14.2truecm
\hskip 7truecm
\includegraphics[scale=0.7]{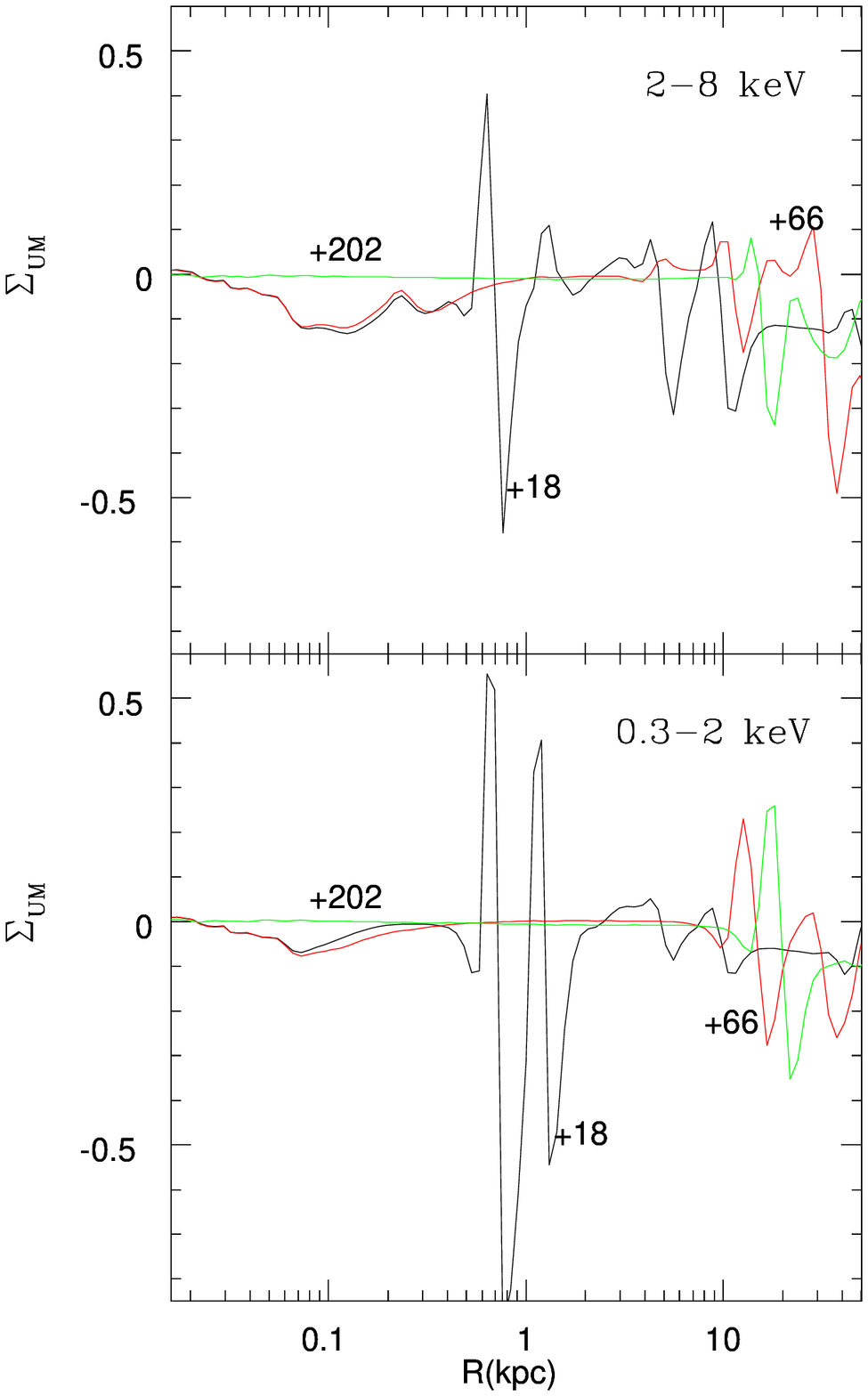}
\caption{Unsharp masked residuals (Eq.~\ref{conv}) for the gas emission of model \bd
 during outburst, at the same times as in Fig.~\ref{f6}. Note the
 ``cavity'' as a decrement in brightness close to the center, at
 $+6$, $+18$ and $+66$ Myr, and the surrounding bright and sharp rim;
both features are similar to what revealed by unsharp masking in a few
well studied ellipticals (e.g., NGC4552, Machacek et al. 2006).}
\label{f12}
\end{figure*}

In analogy with the discussion of the temperature profiles,
Fig.~\ref{f11} shows $\Sigma (R)$ just before, during, and after the
last major nuclear burst at 7.498 Gyr, with times counted from the
first accretion event. At -2 Myr the formation of the off-center cold
shell produces the characteristic feature of a sharp decrease of
$\Sigma$ in the hard band, and a bright rim in the soft band.  The
subsequent curves show the shock moving outwards after the major
burst, and the presence of very hot gas at the center, revealed by the
central peak in the hard band.  The disturbances in the $\Sigma$
profiles due to the shells launched by the repeated sub-bursts are
much more visible in the soft than in the hard band, as particularly
apparent at +18 Myr.  At this time, note the remarkable presence of a
hot center surrounded by a denser and colder shell, producing a sharp
peak in $\Sigma (R)$, at $R=600-700$ pc, and a sharp dip in $\ta(R)$
in Fig.~\ref{f6}.  The final two times (+66 Myr and +202 Myr) show the
result of the degassing caused by the passage of the shock waves: the
gas density is low, and subsonic perturbations remain at a radius of
$\gsim 10$ kpc.  A different representation of the profiles during the
burst phases is given by Fig.~\ref{f12}, where $\Sigma_{\rm UM}(R)$
resulting from the unsharp masking procedure (Eq.~\ref{conv}) is
shown. Fluctuations in brightness that can be clearly distinguished
are that due to the cold shell (time -2 Myr), and, after the burst,
the negative off-center region in $\Sigma_{\rm UM}(R)$ delimited by a
sharp positive peak (times +6, +18, +66 Myr); the latter resembles
what is commonly called a hot gas ``cavity''.

Note that, when implemented in 2D simulations, the same input physics
adopted for the present class of models gives conical hot gas outflows
from the nucleus, during outbursts: hot gas is ejected along the
symmetry axis, so that elongated "cavities" (i.e., regions with a gas
density decrement) are created (Novak et al. 2010). In addition, in
their study of a model very similar to \bt, Jiang et al. (2010) found
after a burst taking place at $6.5$ Gyr, a cavity filled with radio
emitting particles of $\sim 4.4$ kpc in size, detectable during the
first $\sim 10$ Myr after the burst.  Therefore, we expect that
cavities in the hot gas, such as those seen as off-center minima in
the profiles of Figs.~\ref{f11} and~\ref{f12} at times from 6 to 66
Myr after the burst, should be fairly bright in the radio and would
have, in X-rays, two essentially hollow conical lobes. They should
even show non-thermal X-ray emission of the type seen in the Crab
nebula (Jiang et al. 2010). Finally, taking at face value the results
of the present analysis and the estimates on the radio detectability
of Jiang et al. (2010), hot gas cavities seem more long-lasting than
their radio detectability.

\begin{figure}
\vskip -6truecm
\hskip -1truecm
\includegraphics[angle=0.,scale=0.65]{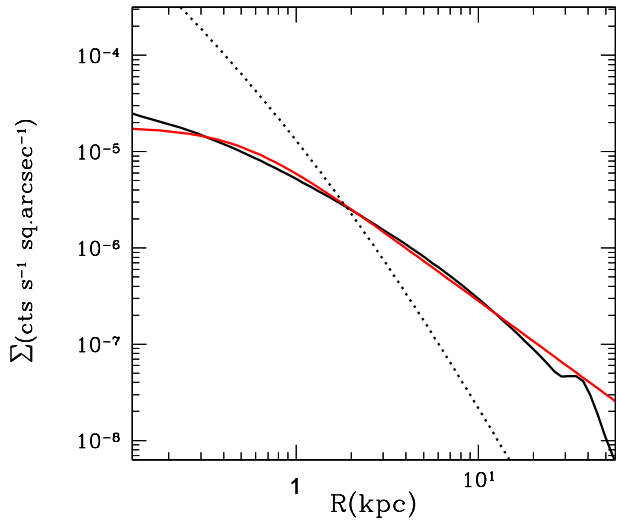}
\caption{The 0.3--2 keV surface brightness profiles of the hot gas for
\bd at an epoch of 9 Gyr (solid black line), compared with
that of a model with the same $\lb$, $\sigma$,
$\re$ and $\lx$ as \bd, but without feedback (dotted line); the conversion from flux to
counts refers to a $Chandra$ ACIS pointing.  Also shown is the best
fit to the hot ISM brightness profile from an $Chandra$ ACIS pointing of the elliptical
NGC4365 (at a distance of 20.4 Mpc), that has $\lb$
and $\lx$ close to that of the models (red, from Sarazin et al.~2003); the
innermost flattening of the red profile within $\sim 200$ pc ($\sim
2^{\prime\prime}$) is due to PSF blurring effects.}
\label{f10}
\end{figure}

Coming to the observability of the predicted features, one major
property of the models is the decrease of $\Sigma (R)$ in the central
galactic regions, produced by the nuclear outbursts, with respect to
models without feedback.  Interestingly, bright ellipticals imaged
with $Chandra$ (e.g., Loewenstein et al. 2001) show a brightness
profile that is quite flat within the central $\sim 1$ kpc, a feature
impossible to reproduce with pure inflow models, while it resembles
the profile of the ``pre-burst'' phase (black lines in Fig.~\ref{f11},
left panels), or at the end of a burst (+202 Myr). To better
illustrate this point, Fig.~\ref{f10} shows the different shape of
$\Sigma(R)$ for \bd and for a model with the same $\lb$, $\sigma$ and
$\re$, but without AGN feedback, and the sole SNIa's heating
included. The two $\Sigma(R)$ profiles considered refer to the
present quiescent epoch, yet the difference in steepness is clear.  
Figure~\ref{f10} also shows the observed $\Sigma(R)$ for an
elliptical at the periphery of the Virgo cluster, with $\lx$ close 
to that of \bd; the agreement between model and observation is very
good.

Another major prediction is given by the disturbances in the profile
during an outburst; these keep above the level of the unresolved
stellar emission, for a deep observation of a nearby galaxy with
$Chandra$, as shown by Fig.~\ref{f11}.  The central spike in $\Sigma
(R)$, during the high-temperature and high-density phase at the
center, is confined within $\sim 100$ pc, and then likely to remain a
central unresolved feature even in galaxies observed at the high
angular resolution of $Chandra$. Disturbances as shells and ripples
farther out in the galaxy  last $\lsim
0.2$ Gyrs, and are more likely to be observed.  Given the typical
duration of these features, and the presence of 3--4 major outbursts
during the whole evolution, statistically they should be present, and
possibly revealed by current X-ray observations, in $\simeq 5-10$\% of
the galaxies with $\lb$ similar to the model galaxy.

In fact, many nearby galaxies show a disturbed appearance, as most
recently revealed by studies based on $Chandra$ data (e.g., Finoguenov
\& Jones 2001, Forman et al. 2005, Soria et al. 2006, Diehl \&
Statler 2007, Nulsen et al. 2009, Baldi et al. 2009, Dunn et
al. 2010).  The ISM morphology of a large set of galaxies (54) shows a
level of disturbance correlated with the radio luminosity derived by
the NRAO VLA Sky Survey (thus including both pointlike and extended
radio emission; Diehl \& Statler 2008a).  Also, many of the best
studied gas-rich galaxies show decrements in the X-ray surface
brightness map, identified as cavities formed when AGN jets inflate
radio lobes and displace surrounding gas; in many cases the cavities
are filled with radio plasma, and surrounded by armlike features,
sometimes classified as shocks. The hot gas disturbances have then
been generally attributed to jet activity.  As
discussed above, we expect that if we took two cones from our
solutions, they would give "lobes". These would produce shocks at the
edges and the lobes would be filled with radio emitting particles.

There are also a few galaxies without currently evident extended radio
emission, but with signs of an outflow and hot central gas, as NGC4552
(Machacek et al.~2006).  This galaxy shows a weak core radio source
unresolved by the VLBA, and in the (unsharp masked) X-ray image two
conspicuous ringlike features at 1.3 kpc from the galaxy center,
surrounding two cavities; these features have been found consistent
with shocked gas driven outward by recent nuclear activity, as in a
bipolar nuclear outflow. The gas temperature in the central $\sim 100$
pc of the galaxy is $1\pm 0.2$ keV, hotter than elsewhere in the
galaxy, suggesting that we may be directly observing the reheating of
the galaxy ISM by the outburst (Machacek et al.~2006).  These
characteristics resemble those predicted by the models for the temperature and
the surface brighntess during the afterburst phase.

\section{Discussion and conclusions}

The hot gas properties of massive ellipticals, with regard to
luminosity and temperature, and their spatial distributions, allow us
to derive insights into the hot gas evolutionary status, and its link
with the host galaxy. Since the cooling times are short compared to
the galaxy age, it is now commonly accepted that repeated cooling
catastrophes have occurred in the past, accompanied by central
starbursts and AGN outbursts.  The interest of a better understanding
of this phenomenon is obvious, as it is directly linked to galaxy
formation and evolution, and to the growth of the central
MBH. Unfortunately, a complete theoretical picture is still missing, so that
modeling coupled with a close comparison with observations is crucial
in this field. In fact, these feedback events must leave signatures on
the X-ray properties of the galaxies; indeed, the observed temperature
and brightness profiles often cannot be fit easily with smooth
profiles, or cooling flow profiles (e.g., Sarazin 2011, Statler 2011).
In this work we have calculated the observational properties in the
X-ray band of two galaxy models, representative of a class of detailed
simulations of physically based models for the investigation of
feedback modulated accretion in isolated galaxies. The feedback
physics includes the combined effects of radiation and AGN winds
(Paper III).  The observational properties derived provide good
matches to what observed in general for the local universe, and
account for a few otherwise puzzling observed properties; on the other
hand, they also evidence the need for changes or additions to the
input physics.  The main results of the present investigation are the
following:

1) After an evolution of $\simeq 10$ Gyr, the models are typically in a
permanent quiescent phase.  The bolometric nuclear emission is very
sub-Eddington ($l \simeq 10^{-4}$), within the range observed for $l$,
though the most frequently observed values ($l\approx 10^{-5}$ or
less) are somewhat lower. Unfortunately, uncertainties in the
bolometric correction to apply to observed nuclear luminosities,
appropriate for a spectral energy distribution at low emission levels,
do not allow to make a stringent comparison between modelled and
observed values.  However, also the nuclear X-ray emission $L_{\rm
BH,X}$ of the models, estimated as $L_{\rm BH,X} \lsim 0.2\lbh$ as
should be appropriate for low luminosity nuclei, and $L_{\rm
BH,X}/\ledd$, tend to lie on the upper end of what is observed.  Thus
in real galaxies an additional mechanism may be required to reduce
further the mass available for accretion; this could be provided by
the mechanical feedback of a (nuclear) jet, and/or by a thermally
driven wind from a RIAF. Alternatively, the switch from disc to RIAF
behavior (that is $l\propto \dot m^2$) should occur at a larger $l$
than assumed here (e.g., at $l\simeq 0.1$ rather than $l\simeq 0.01$).
Simulations of ram-pressure stripping effects, instead, showed that a
reduction in the accretion rate is not attained, because in the quiescent
hot accretion regime, the accreting mass comes mainly from the
stellar mass losses within the central $\sim 100$ parsecs of the galaxy,
that are not affected by stripping (Shin et al. 2010b).

2) The X-ray luminosity of the ISM oscillates in phase with the
nuclear luminosity, though with much broader peaks; at the present
epoch, $\lx$ lies at the lower end of the large observed range for
galaxies of $\lb$ similar to that of the models.  The
degassing/heating in the models may then be too efficient, or a
larger/more concentrated gravitating mass, or a confining external
medium, are needed.  However, when the gas luminosities during the
whole evolution are considered, the observed $\lx$ range is better
reproduced.  This is even more true if also an additional model of
larger $\sigma$, and same $\lb$, is included.  Part of the observed
large variation in $\lx$ for galaxies of a given $\lb$ could then be
explained by nuclear activity.

3) The average ISM temperature is within the observed large range for
the model $\sigma$; when estimated within $\re$, the model
temperatures reproduce better the upper half of those observed.
Modifications to the models by the addition of a jet or an external
medium, as suggested in the previous points, should then not increase
the average temperature within $\re$.

4) During quiescence, the profiles of the gas temperature and
brightness resemble those observed for many local galaxies. Especially
remarkable is the lack of the steep brightness profile shape typical
of inflowing models, due to the frequent removal of gas from the
galactic central regions, and to the heating provided by mechanical
feedback (that is always present, even during quiescent phases).  The
models show negative temperature gradients, that are common for
isolated galaxies; the addition of a jet or a confining agent should
change the temperature profiles into a flat or outwardly increasing
profile, as also frequently observed for galaxies in rich
environments.

5) During outbursts, disturbances are predicted in the temperature and
brightness profiles; the ISM resumes the smooth appearance of steady
and low-luminosity hot accretion on the MBH on a time-scale of $\lsim
200$ Myrs.  The most conspicuous variations with respect to smooth
profiles are within current detection capabilities, and could
correspond to (part of) the widespread disturbances observed in
galaxies of the local Universe. In particular, shocked hot gas should
be seen at the galactic center (within $\sim 100$ pc), possibly not
resolved; this would be a certain sign of prior AGN activity.  These
hot bubbles could be revealed by emission of
cosmic-rays, in a structure similar to a gigantic supernova remnant.
Preliminary results from 2D runs show bipolar nuclear outflows, that
should be seen as conical cavities extending from the galactic center,
and may be called jet-like features.

6) The duty-cycle of nuclear activity is of the order of a few $\times
(10^{-3}-10^{-2})$, depending on the assumed mechanical feedback
efficiency; in general, a burst cycle lasts for $\approx 10^7$ yrs.
These duty-cycle values are broadly consistent with the fraction of
active galaxies measured in observational works, though reported
values for the local universe are somewhat lower, for the MBH mass of
the models.  In order to make a more consistent comparison with
observations, the dataset of models should be increased, and the duty
cycle computed only for the last 2--3 Gyrs; this will reduce the duty
cycle, as the models are characterized by a declining nuclear
activity.

7) The duty-cycle of perturbances in the ISM is of the order of
5-10\%, from their average number and duration.  This duty-cycle
likely increases with galaxy mass, because an outburst has a greater
impact in less massive (and less gas-rich) systems, which then are
"on" for a shorter time (Ciotti \& Ostriker 2011). The ISM duty-cycle,
and its trend with galaxy mass, compare reasonably well with
preliminary estimates obtained from a large sample of hot gas coronae
in elliptical galaxies observed with $Chandra$ (Nulsen et al. 2009):
the fraction of galaxies with X-ray cavities in the hot gas is $\lsim
10$\% when $\lx<10^{41}$ erg s$^{-1}$ (as for the models), and reaches
$\sim 25$\% in the most luminous ones. The presence of cavities has
been attributed to the action of jets inflating radio lobes and
displacing the surrounding gas. Cavities can also be created in the
scenario presented here, as hinted for by 2D simulations, due to
bipolar nuclear outflows.

7) Two diagnostic planes have been constructed. In the first one, the
nuclear luminosity $\lbh$ and the ISM luminosity $\lx$ are followed
during the whole model evolution. The points representative of the
models populate a wedge region, that should then be occupied when
observing a large set of galaxies.  Another plane shows the evolution
of $\lx$ versus the average gas temperature $\ta$;
here the most populated region is that of a large $\lx$ variation
(factor of $\sim 10$) for $\ta$ keeping between 0.4 and 0.6 keV.

Clearly, a larger set of models is to be explored, in order to better
establish the final gas properties (as gas content, nuclear and ISM
duty-cycle, etc.) to be compared with those of a statistically large
sample.  For example, a general expectation is that changes of the
galaxy properties have an impact on the number of nuclear outbursts:
depending on many model parameters (supernova rate, central $\sigma $,
dark matter amount and distribution, and even external pressure due to
an intragroup or intracluster medium), bursts could take place even
towards the present epoch, or be confined to the early epoch (Ciotti
\& Ostriker 2011).

\acknowledgments
L.C. and S.P. are supported by the grant MIUR PRIN2008

%%%%%%%%%%%%%%%%%%%%%%%%%%%%%%%%%%%%%%%%%%%%%%%%%%%%%%%%%%%%%%%%%%%%%%%%%%%

\clearpage

%%%%%%%%%%%%%%%%%%%%%%%%%%%%%

\begin{table*}\footnotesize
%\tabletypesize{\scriptsize}
\caption[] {The representative models}\label{modref}
\begin{tabular}{lcccccccc}
\noalign{\smallskip}
\hline
\noalign{\smallskip}
 Model&  $\epswM$& $<\epsw>$& $<\eps_{\rm EM}>$&  $\log \Delta\mbh$&
 $\log \Delta M_*$ &  $\log \Delta M_{\rm w}$&  $\log \mgas$ &  $\log
 \lbhefopt/\ledd$ \\ 
(1)&
 (2)&
 (3)&
 (4)& 
 (5)&
 (6)&
 (7)&
 (8)&
 (9) \\
\noalign{\smallskip}
\hline
\noalign{\smallskip}
{\bf \bd}  & $10^{-3}$        & $2.0\,10^{-5}$              & 0.105  & 8.74 & 9.74  & 10.27 & 9.68 & -5.13\\
{\bf \bt}  & $3\times 10^{-4}$& $1.2\times 10^{-5}$ & 0.133  & 9.05 &
10.22 & 10.31 & 9.34 & -5.43\\
\hline 
\end{tabular}
\smallskip
\tablecomments{Relevant model properties at an age of 12 Gyrs; masses
 are in units of solar masses and luminosities in erg s$^{-1}$. The
 value of $\epswM$ is reached when $\lbh\geq 2\ledd$, and the maximum
 radiative efficiency is set to $0.2$.  Columns 3 and 4 give the
 accretion weighted values of the mechanical and radiative efficiencies.
 $ \Delta\mbh$ is the total accreted MBH mass,
 $\Delta\mast$ is the total stellar mass formed during the
 evolution, $\Delta M_{\rm w}$ is the total amount of ISM lost at
 10$\re$, and $\mgas$ is the amount of gas inside $10\re$.
 $\lbhefopt$ is the fiducial MBH luminosity in the optical as would
 be seen at infinity after absorption, with $\lbhefopt=0.1\lbh$ at
 the first grid point (see Paper III for details). }
\end{table*} 

%%%%%%%%%%%%%%%%%%%%%%%%%%%%%%%%%%%%%%

\begin{table*}\footnotesize
\caption[] {Nuclear and gas emission properties (at 12 Gyr)}\label{tab2}
\begin{tabular}{ l  c c c c c c c c c }
\noalign{\smallskip}
\hline
\noalign{\smallskip}
Model   &  $\mbh$  & log$\lbh $          & $l$ & $L_{\rm BH,X} $ &$L_{\rm BH,X}/L_{Edd}$ &  &duty cycle &  &$\log L_{\rm X} $ \\
           &($M_{\odot}$)&(erg s$^{-1}$) &  ($10^{-4}$)     & (erg s$^{-1}$) &                                & Bol & Opt & UV & (erg s$^{-1}$)\\
 (1)      &   (2)          &         (3)          & (4)  &     (5)                  &                     (6)       & (7)  & (7) & (7)  & (8)\\
\noalign{\smallskip}
\hline
\noalign{\smallskip}
{\bf \bd }&   $8.4\times 10^8$ & 43.39  & 2.0
& $\lsim 5\times 10^{42}$  & $\lsim 2\times 10^{-5}$& $6.3\times
10^{-3}$ &$3.2\times 10^{-3}$ &$3.0\times 10^{-3}$ & 40.1 \\
{\bf \bt  }&   $1.4\times 10^9$ & 43.38& 1.0
& $\lsim 5\times 10^{42}$  & $\lsim 2\times 10^{-5}$& 
$4.8\times 10^{-2}$ & $1.8\times 10^{-2}$ & $8.6\times 10^{-3}$ &39.6 \\
\hline
\end{tabular}
\smallskip
\tablecomments{Column (1): galaxy model; col. (2): final MBH mass; cols. (3) and (4):
bolometric nuclear luminosity and its Eddington ratio, for $A=100$ in Eq.~\ref{effic};
cols. (5) and (6): 0.3--10 keV nuclear luminosity and its Eddington ratio, for
radiatively inefficient accretion (see Sect.~\ref{nuc}); col. (7): the
duty cycle calculated over a temporal baseline of 6--13 Gyr, in the
bolometric, optical, and UV bands (see Sect.~\ref{nuc} for more details); col. (8):
the 0.3--2 keV gas luminosity within $10\re$, at 12 Gyr.}
\end{table*}

\clearpage

%%%%%%%%%%%%%%%%%%%%%%%%%%%%%%%%%%% FIG1
%% Fig. 1 :
%%%%%%%%%%%%%%%%%%%%%%%%%%%%%%%%%%%%%%

\clearpage

%%%%%%%%%%%%%%%%%%%%%%%%%%%%%%%%% FIG 2a-2b
%%%%%%%%%%%%%%%%%%%%%%%%%%%%%%%%%%%%%%%%%%%%%%%%

\clearpage

%%%%%%%%%%%%%%%%%%%%%%%%%%%%%%%%%%%%%%%%%%%% FIG 3

\clearpage

%%%%%%%%%%%%%%%%%%%%%%%%%%%%%%%%%%%% FIG 4
%%%%%%%%%%%%%%%%%%%%%%%%%%%%%%%%%%%%%%%%%%%%%%

\clearpage

%%%%%%%%%%%%%%%%%%%%%%%%%%%%%%%%%%%% FIG 5
%%%%%%%%%%%%%%%%%%%%%%%%%%%%%%%%%%%%%%%%%%%%%%

\clearpage

%%%%%%%%%%%%%%%%%%%%%%%%%%%%%%%%%%%%%%%%%%%%%% FIG 6

\clearpage

%%%%%%%%%%%%%%%%%%% T profiles in quiescence: FIG 7a, 7b
%%%%%%%%%%%%%%%%%%%

\clearpage

%%%%%%%%%%%%%%%%%%% FIG 8 a,b,c,d 
%%%%%%%%%%%%%%%%%%%%%%%%%%%%%%%%%%%%%%%%%%%%%%%%%%%%%%%%%%%

\clearpage

%%%%%%%%%%%%%%%%%%%%%%%%% FIG 9

\clearpage

%%%%%%%%%%%%%%%%%%%%% FIG 10
%%%%%%%%%%%%%%%

\clearpage

%%%%%%%%%%%%%%% FIG 11
%%%%%%%%%%%%%%%%%%%%%%%%%%%%%%%%%%%%%%%%%%%%%

\clearpage

%%%%%%%%%%%      FIG 12
%%%%%%%%%%%%%%%%%%%%%%%%%%%%%%%%%

\end{document}